%% file: PSACCF.tex
\documentclass[lettersize,journal]{IEEEtran}
\usepackage{amsmath,amsfonts}
\usepackage{amsthm}
\usepackage{amssymb}
\usepackage[linesnumbered, ruled, vlined]{algorithm2e}
\usepackage{array}
\usepackage[caption=false,font=footnotesize,labelfont=rm,textfont=rm]{subfig}
\usepackage{textcomp}
\usepackage{stfloats}
\usepackage{url}
\usepackage{verbatim}
\usepackage{graphicx}
\usepackage{cite}
\usepackage{color}
\usepackage[hidelinks]{hyperref}
\graphicspath{{figs/}}
\begin{document}

\title{PSACCF: Prioritized Online Slice Admission Control Considering Fairness in 5G/B5G Networks}

\author{Miao Dai, \textit{Member, IEEE}, Long Luo, \textit{Member, IEEE}, Jing Ren, \textit{Member, IEEE}, \\ Hongfang Yu, \textit{Member, IEEE}, Gang Sun, \textit{Member, IEEE}
%	IEEE Publication Technology,~\IEEEmembership{Staff,~IEEE,}
%        % <-this % stops a space
\thanks{Miao Dai is with Key Lab of Optical Fiber Sensing and Communications (Ministry of Education), University of Electronic Science and Technology of China, Chengdu, China (e-mail: daimiao@std.uestc.edu.cn).}% <-this % stops a space
\thanks{Long Luo and Jing Ren are with Key Lab of Optical Fiber Sensing and Communications (Ministry of Education), University of Electronic Science and Technology of China, Chengdu, China (e-mail: longluo.uestc@gmail.com; renjing@uestc.edu.cn).}
\thanks{Hongfang Yu is with Key Lab of Optical Fiber Sensing and Communications (Ministry of Education), University of Electronic Science and Technology of China, Chengdu, China; and is also with Peng Cheng Laboratory, Shenzhen, China (e-mail: yuhf@uestc.edu.cn).}
\thanks{Gang Sun is with Key Lab of Optical Fiber Sensing and Communications (Ministry of Education), University of Electronic Science and Technology of China, Chengdu, China; and he is also with Agile and Intelligent Computing Key Laboratory of Sichuan Province, Chengdu, China (e-mail: gangsun@uestc.edu.cn).}
%\thanks{Manuscript received April 19, 2021; revised August 16, 2021.}
}



\maketitle

\input{section/sec00.Abstract}

\input{section/sec01.Introduction}

\input{section/sec02.RelatedWorks}

\input{section/sec03.PreliminaryInformation}

\input{section/sec04.SystemModel}

\input{section/sec05.AlgorithmDesign}

\input{section/sec06.PerformanceEvaluation}

\input{section/sec07.ConclusionFutureWork}

\bibliography{refs}

\end{document}

%% file: section/sec00.Abstract.tex
\begin{abstract}

\color{black}5G/B5G is envisioned to support various services with the assistance of network slices, each slice instance asks for adequate resources to provide the pre-negotiated service quality to its subscribers. Slice Admission Control (SAC) algorithm is a necessity for Slice Providers (SPs) to guarantee the QoS and QoE of each admitted request with limited resources. In that circumstance, the priority concern of services and the fairness of resource allocation arise as meaningful topics for researchers. The former originates from the innate characteristics of various services supported by 5G networks, and the latter matters because slices are instantiated on shared physical equipment. However, the two issues are mainly investigated separately in the literature or do not receive sufficient research simultaneously. In this work, we study the SAC problem in 5G/B5G networks, aiming at enhancing the fairness degree on the premise of satisfying the necessary priority requirements. We first reinterpret priority as a higher cumulative service acceptance ratio (CSAR), and adopt the uniformity of adjacent CSAR gaps to reflect the fairness. Based on these adjustments, the SAC problem is formulated as a non-linear and non-convex multi-objective optimization. Thus, we propose a heuristic algorithm called Prioritized Slice Admission Control Considering Fairness (PSACCF) to solve it. It introduces the resource efficiency of services to amend priority violations, then promotes fairness by setting the target CSARs for each service type and pushing their actual CSARs toward. Numerous simulations are carried out to compare the performance of PSACCF with two existing algorithms, termed MHPF and AHPF. Results show that our algorithm can achieve a nearly identical priority indicator to the comparisons, as well as at least a 33.6\% improvement in fairness degree and a higher minimum average resource utilization.

\end{abstract}


%% file: section/sec01.Introduction.tex
\section{Introduction and Motivations} \label{introduction}

\IEEEPARstart{F}{uture} applications are believed to belong to three broad categories~\cite{shafi20175g}: enhanced Mobile Broadband (eMBB), ultra-Reliable Low Latency Communication (uRLLC), and massive Machine Type Communication (mMTC), where high mobility of devices and large communication bandwidth, strict time delay and reliability, sufficient access capacity should be satisfied respectively. They focus on different performance indicators, and the type and amount of resources consumed are also distinct. To handle the heterogeneity of these applications, network slice~\cite{chahbar2020comprehensive,zhang2019overview} is employed by 5G/B5G networks to realize customized services. It brings a new paradigm that allows to deploy virtual networks upon generic equipment. Through designing and adjusting the structure of slices, Slice Providers (SPs) can establish logic networks that possess corresponding abilities required by Slice Tenants (STs).

In such a circumstance, slices rather than tailored devices are the basic units to accommodate service requests, which makes the design, deployment, and maintenance of services more flexible and efficient. However, slice just provides a template of networks, only when enough resources are allocated, i.e., instantiated, can it effectively plays the role STs desire. Since the physical resources are limited, admission control is needed to match the amount of available physical resources and the admitted workload.

Admission control is a policy deciding whether to accept or reject a perceived service request when the total demand is unaffordable, which had attracted the attention of researchers since the 1980s. In the early studies~\cite{posner1985traffic,ramjee1997optimal}, requests remain homogeneous, only one kind of resource demand is considered and the optimization objective is also unsophisticated. Nevertheless, with the continuous development of applications, the resource requirements of services are diversified, the heterogeneity between services is strengthened and the indicators concerned are different. Thus, in sliced networks, call admission control turns into slice admission control, which is expected to undertake a more complex task.

In 5G/B5G networks, many factors should or can be considered as objectives of SAC algorithms~\cite{ojijo2020survey}, mainly are resource constraint, the priority of services, fairness of resource sharing, and profits earned~\cite{tang2019service}. Resource constraint is the most basic target and must be met because each slice is attached to a Service Level Agreement (SLA) which can only be guaranteed when obtained the necessary amount of resources~\cite{li2021automated,nourian2021practical}, the risk of destruction in isolation between slices will rise if resource squeeze happens. What's more, with the evolution of applications, the dimension in resource demand also expands. For example, model training in machine learning consumes plenty of computation and memory resources, services providing file download need large disk space for data storage, and video and live streams occupy bandwidth resources to transmit traffic. Some new applications even ask for all these concurrently, such as cloud games~\cite{zhang2020discussion}, VR, and AR~\cite{siriwardhana2021survey}, where a lot of computing power, memory capacity, and bandwidth are spent to track and process user instructions, render the screen, and send back the pictures. Due to the reasons mentioned above, more rigorous and precise resource constraints should be imposed on SAC algorithms.

Priority, an attribute of services, also received concerns in some admission works. There are many reasons to make it a significant issue. From the perspective of users' quality of experience (QoE), handoff requests due to mobility should be privileged over new requests, because interrupting an ongoing service is believed more disturbing than rejecting a new one. Besides, services with distinct degrees of urgency ought to be treated differently, it's reasonable to set higher priority to requests for emergency communication than those for general applications. Sometimes, profit is also a vital cause of priority. Since each kind of request is served by the corresponding slice, the hierarchical relationship among services is naturally inherited by slices and remains as an optional objective of SAC algorithms. Expect for priority, fairness is another interesting topic in situations where multiple kinds of services coexist and share the common substrate networks. Optimization for fairness can achieve more balanced resource allocation and ensure that each kind of request has the opportunity to be served. However, the concept of fairness was traditionally equivalent to that of average and equality~\cite{hwang2005call}, which seems inappropriate~\cite{ogryczak2014fair} if the priority is needed as a premise.

In this paper, we focus on the slice admission control problem in 5G/B5G networks, and manage to develop a SAC algorithm where priority is taken as the first objective and fairness is put under control simultaneously, which is rarely investigated before. The main contributions of this paper are summarized below:

\begin{itemize}
	\item Newly defined concepts about priority and fairness are proposed to let the two traditionally contradictory objectives compatible, making them more suitable to the environment where heterogeneous applications coexist on the shared infrastructure. Thus, leading it a reasonable and practical thing to consider them concurrently in the process of slice admission control.
	
	\item A multi-queue-based online SAC algorithm (called PSACCF) that can distinguish the priority among service requests and maintain necessary fairness on top of that is designed. Two parameters are used to control the degree of priority between adjacent priority levels and the threshold of acceptable fairness respectively. By adjusting these two parameters, the effect of dynamic resource reservation for high priority requests can be realized, which makes our algorithm more flexible when facing different target preferences.
	
	\item Impatient behaviors of service subscribers are taken into account by PSACCF. Rational users are assumed to balk with the probability proportional to the queue length upon arrival or renege when waiting too long in the queue. PSACCF can sense these behaviors that happened inside each queue and react accordingly to hold a stable performance.
	
	\item A lot of simulations are carried out to demonstrate the advantages of PSACCF over some existing works.
\end{itemize}

The remainder of this paper is organized as follows. In the next section, we introduce some relevant works conducted by other researchers. Our system model and concept definitions are given in Section~\ref{system model}, and the details of our slice admission control algorithm is elaborated subsequently in Section~\ref{algorithm design}. Simulation results and data analysis are presented in Section~\ref{performance evaluation}. Finally, Section~\ref{conclusion} concludes this paper.

%% file: section/sec02.RelatedWorks.tex
\section{Related Work}

Tremendous efforts have been made to admission control in recent decades, brief reviews about them are listed in this part to explain their scope of applications and limitations.

Profit is an ordinary objective favored by many researchers. Authors in~\cite{bakri2021using} proposed several methods based on reinforcement learning to maximize the revenue of the infrastructure provider, aforementioned three categories are taken into account but only one request arrives each time and a single kind of resource is considered. A similar optimization target and method are adopted in~\cite{villota2021admission}, and the processing capability of nodes is added to resource constraints. Paper~\cite{challa2019network} concentrates on the tradeoff between revenue and service admission rate, which was formulated as a multiple knapsack problem. To reduce the complexity, all requests are roughly divided into premium or best-effort, and an adjustable quota vector is allocated to each kind to control the tradeoff. Paper~\cite{ebrahimi2020joint} did the same thing but from a different angle, it managed to minimize the network cost of the service provider at a premise that the predefined criteria of requests are met. Computing, memory, and storage capacities are concerned, and each time the resource constraints are not held, the requests demanding more resources are iteratively rejected until the solution becomes feasible. Authors in~\cite{han2020multiservice} proposed a multi-queue-based admission control model to show its advantage in handling heterogeneous tenant requests. Rational behaviors are concerned, under which higher profit is desired, a method based on queuing theory is used to analyze the response of queue states to admission results. Nevertheless, the access preference list was just randomly generated there, and no deterministic calculation approach is given.

As new applications continue to develop and evolve, they are deeply refined and reclassified on top of the three broad categories, which makes the priority issue non-negligible. In~\cite{han2018admission}, two kinds of slices are defined, guaranteed service and best-effort service, the former is believed to be more important and should be preferred when conducting admission control. To realize it, a Q-learning approach is proposed, where higher rewards are attached to the admission of guaranteed services, and a negative reward is obtained when dropping such a request. Instead of assigning priorities to different services, authors in~\cite{monemi2015low} put them to each wireless network tier, users in some tiers are prior to those in other tiers. Channel allocation and SINR control are the main focus of this paper to maintain the priority constraint, which means that if a user is provided with its target SINR then those more privileged than him should also get their target SINR. Papers~\cite{caballero2018network,caballero2019network} investigate the admission control and resource allocation model in multi-tenant mobile networks with strategic tenants, a game theory method is employed to maximize the overall utility, and the priority difference of users is indirectly reflected in the multiplier before their respective utilities. Scenario in~\cite{al2020priority} changes to vehicular networks, where various vehicular applications with different priorities coexist. The BS association policy and bandwidth reservation scheme are designed with priority concerns, by assigning higher weights for those more preferred vehicles in their achievable data rates that constitute the final optimization objective.

Fairness is always an important issue when services are provisioned on shared infrastructures. However, it's a controversial topic about what is fairness, in some early studies, it refers to the concept of equality. Authors in~\cite{hwang2003fairness,hwang2005call} investigated the call admission control policy with equal call blocking probability as the fairness measurement. Paper~\cite{yang2020genetic} pay attention to the spectrum allocation across slices in the radio access networks, the fairness here is evaluated by checking the number of qualified users that obtain required bitrates in each slice, higher degree of fairness is achieved when numbers approach each other, a genetic algorithm is designed to perform the allocation process. While this interpretation is considered to make less sense when different applications have distinct QoS requirements. In~\cite{dianati2005new}, a utility-based fairness index is defined, and bandwidth resources are shared among single-hop wireless nodes to achieve near the same utility on each wireless connection. Authors in~\cite{caballero2017multi} think only the contributions of each tenant are taken into account, can it turns out to be fair. So their contributions are treated as a reference when allocating the shares of resources in base stations.

Except for the literature introduced above where priority and fairness are optimized individually, papers~\cite{nasser2004optimal,liu2014fairness} put them together and produce a novel and complex admission problem. In~\cite{nasser2004optimal}, the handoff dropping probability is used as the indicator of QoS, services with higher priorities should be protected with better acceptable QoS to show the preferential treatment, and the gap of dropping probabilities between two neighbor priorities is constrained to a constant threshold to promise fairness meanwhile. Similarly, in~\cite{liu2014fairness}, it's call blocking probability that acts as the criteria to differentiate priorities, handoff calls should experience lower blocking probability than new calls. While fairness is treated differently under two subcases, requests belonging to the same kind of service should have similar blocking probability in different end devices, and requests for different services from the same device ought to obey a predefined ratio in their average acceptance probability. Markov decision process is adopted by both of the two papers to represent the admission policy. Therefore, traffic patterns are usually required as known conditions or can be derived from historical information by them.

Although there have been studies about admission control issues as illustrated above, they have obvious limitations, such as only a single kind of resource is considered, priority and fairness are rarely handled simultaneously, requests with different priorities are actually with homogeneous objectives, i.e., their heterogeneity is not so drastic, and rational behaviors of tenants are not confronted. Hence, there still remains so much work to do on SAC algorithms.

%% file: section/sec03.PreliminaryInformation.tex
\section{Preliminary Information}

In this section, we plan to give essential descriptions of the scenario characterization and concept definition employed in our work.

\subsection{Scenario characterization}

We introduce a general description of the scenario in which this work is carried out. Two main aspects are included, i.e., the characteristics of services and the impatient behaviors of service subscribers.

Except for the aforementioned service classification criterion, the severity of resource requirement can act as another one to further divide services into two classes, known as elastic and inelastic service~\cite{chiu2005network}:

\begin{itemize}
	\item \textbf{Elastic service}. These services do not have strict real-time resource requirements, but ask for a minimum average resource allocation instead. They usually carry an explicit lifetime duration and the total amount of data to transmit or calculate before expiration. Temporary resource shortage will not lead to service interruption as long as it is made up within the subsequent lifetime. Besides, the provision of additional resources can accelerate the completion of services.
	
	\item \textbf{Inelastic service}. On the contrary, inelastic services have stringent resource requirements all the time, meaning that an inadequate supply of resources will affect the QoE terribly or even corrupt the normal operation of services. Sometimes, inelastic services do not carry a definite lifetime duration and can terminate at any time point according to the wishes of users, the service provider cannot know this in advance. {\color{black} For this kind of service, resource over-provisioning can neither make an obvious improvement of QoE nor stimulate the completion of services, but lead to a downside of resource utilization.}
\end{itemize}

%Elasticity reflects the different characteristics of services in resource demand, it's also an explanation of SLA assurance from the perspective of resources. On one hand, elasticity puts fundamental resource constraints on admission decisions, on the other hand, the admission process can become more flexible by fully exploiting service elasticity.

\begin{table}[!h]
	\caption{Application Classify} \label{tab:classify}
	\centering
	\renewcommand{\arraystretch}{1.5}
	\begin{tabular}{lcc}
		\hline
		Applications & By performance & By resource \\
		\hline
		Video call & eMBB & Inelastic \\
		Live stream & eMBB & Inelastic \\
		4K/8K video & eMBB & Elastic \\
		Large file transfer & eMBB & Elastic \\
		
		Smart grid & uRLLC & Inelastic \\
		Automatic driving & uRLLC & Inelastic \\
		
		Smart home/city & mMTC & Elastic \\
		Environmental monitoring & mMTC & Elastic \\
		
		Industrial control & mMTC, uRLLC & Inelastic \\
		Remote driving & eMBB, uRLLC & Inelastic \\
		Remote healthcare & eMBB, uRLLC & Inelastic \\
		Cloud game/VR/AR & eMBB, uRLLC & Inelastic \\
		\hline		
	\end{tabular}
\end{table}

{\color{black}
Several typical applications of 5G networks are listed in Tab.~\ref{tab:classify}, together with their categories based on two metrics, i.e., the performance requirement and resource strictness. Notice that some applications can occupy more than one class under the former criteria, because of their rigorous requirements on multiple performance targets. And those belonging to the same kind among eMBB, uRLLC, and mMTC may continue to be divided into opposite groups according to the stringency of their resource requirements. 
}

The second thing worth mention is the behavior of service subscribers, who are assumed rational here. Sometimes, requests cannot be admitted immediately and queues are formed due to resource unavailability. Thus, some subscribers may revoke the proposed requests or just not submit a request to this overloaded slice provider. We call these impatient behaviors and two cases are considered here~\cite{wang2010queueing}:

\begin{itemize}
	\item \textbf{Renege}. This describes the behavior that a service subscriber cancels his request which is already in queue when it is not pulled to service for too long a time after entering the queue. It can be realized by setting a hold time value to each request, it is invisible to the slice provider and begins to decrease once the request join a queue. As soon as this value saturates, the request departures from the queue.
	
	\item \textbf{Balk}. When a subscriber generates a request, he may not send it to the slice provider if there already exists many requests in the queue of its type. And the longer the length, the lower the willingness of the request to join the queue. It can be interpreted that a long queue means the SP is overwhelmed because of too many requests or this kind of request is just not favored by the SAC algorithm. Thus, the waiting time may exceed its hold time value, leading to a high risk of renege if enters the queue. This behavior can be depicted by introducing a balk probability, which decreases with the increase of queue length.
\end{itemize}

%These two behaviors are common in reality. The purpose of incorporating them into the premise hypothesis is to maintain the accuracy and authenticity of the scenario description, so as to design a slice admission control algorithm that can be applied in practice.

\subsection{Adjustment of concepts}

In this subsection, we concentrate on the characterization of priority and definition of fairness employed in this paper. First, the reasons why the traditional definitions are no longer appropriate are presented, and then new interpretations suitable to the SlaaS environment are given.

\subsubsection{Characterization of priority} \ \par

The difficulty lies in how to characterize priority concretely, so as to judge whether priority is guaranteed by SAC algorithm. In some previous studies~\cite{challa2019network, li2021research}, priority is expressed by imposing additional exclusive resources on those with higher level. Although these methods ensure the priority difference between services to a certain extent, it often only supports single kind of resource, so cannot migrate directly to the scenarios where multiple kinds of resources are considered and different services even rely on diverse resources with different degrees. Assigning weights is also a common method of maintaining priority. By giving higher weight to the high priority service in the optimization goal, the admission algorithm can be prompted to access more high priority services. However, this usually requires a homogeneous target among prioritized services, which is not always held here.

In order to overcome the difficulty of service heterogeneity when characterizing priorities, the indicator used to measure must be service independent. Thus, we adopt the cumulative service acceptance ratio (CSAR), which is complementary to the service blocking ratio in~\cite{ramjee1997optimal}, to differentiate the priority. If the size relationships of CSARs are consistent with that of priority levels, it's considered that the priority difference is guaranteed. In this way, no matter which kind of resources the services mainly consume or what performance indicators they are concern about mostly, the priority relationships among them can be explicitly evaluated.

Compared with the above methods, using CSAR to characterize priority will neither require a smart quota distribution of resource capacity, nor bring drastic service discrimination, because the concept of priority no longer equals absolute privilege, but preference instead. That means low priority services will not always be ignored even when they coexist with high priority ones. What's more, by adjusting the gap between the CSARs of adjacent priority services, the degree of priority can be put under control conveniently, which allows us to perform a more precise and flexible admission decision.

\subsubsection{Definition of fairness} \ \par

Fairness has always been an enduring topic, the level of fairness is believed to impact the performance of networks~\cite{haryadi2017fairness} and QoE of services~\cite{perveen2021dynamic}. However, it's never a well-investigated problem, because the definition of fairness is extremely scenario-dependent. And optimizing fairness is usually opposite to other objectives pursued simultaneously, like the priority requirement. Therefore, a reasonable definition of fairness must be provided in advance.

The traditional concept of fairness borders on that of average and equality. Unfortunately, it's increasingly partial to treat fairness as equality~\cite{ogryczak2014fair,yang2020genetic}. Even for homogeneous services, admission can only be considered fair if the respective contributions of each kind are taken into account~\cite{caballero2017multi}. Let alone when different services have heterogeneous resource demands, diverse performance preferences, and distinct priority levels. Giving a preference to particular services based on their characteristics happens to be fair in the latter case. Evidence can be found in articles~\cite{nasser2004optimal,liu2014fairness}, where the phenomenon that high priority service deserves lower blocking probability is taken for granted. Both of them use a threshold to control the fairness there. More specifically, ~\cite{nasser2004optimal} sets the threshold to the absolute gap between adjacent dropping probability, while~\cite{liu2014fairness} sets the threshold to the ratio of adjacent average acceptance probability.

Inspired by these works, we place the embodiment of fairness on the uniformity of gaps in CSAR between adjacent priority levels after zooming with their relative weights. In other words, fairness reaches its peak when the admission result produces equal zoomed CSAR gaps instead of the original CSARs. Under this definition, fairness can be improved without sacrificing the degree of priority, and information from all adjacent priority pairs is integrated to make the measurement of fairness more balanced and comprehensive.

%% file: section/sec04.SystemModel.tex
\section{System Model} \label{system model}

In this section, we elaborate the system model with mathematical symbols to make more rigorous and explicit expressions.

\subsection{Overall architecture}

\begin{figure}[!h]
	\centering
	\includegraphics[scale=0.5]{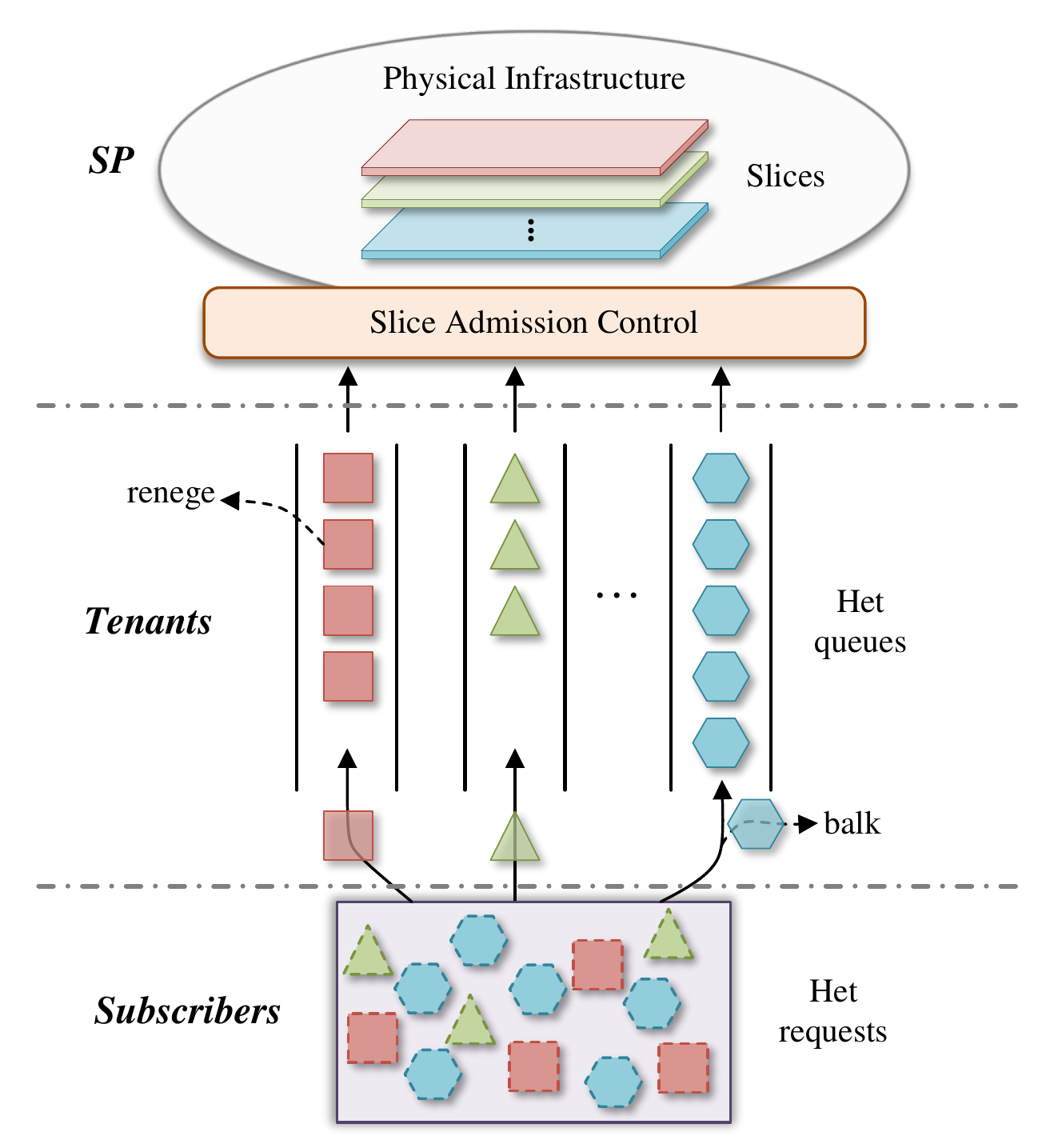}
	\caption{System architecture}
	\label{fig:architecture}
\end{figure}

In general, the infrastructure owners can act as slice providers to offer virtual networks established on their physical equipment. We take a more generic case where each SP can only rent his infrastructure to tenants in terms of virtual slices, because they have no right to schedule resources belonging to other InPs without permission. If several SPs negotiate to interoperate resources for special gains like a wider geographical coverage of services, they can be regarded as one united SP whose resource capacities are equal to the sum of each member. \textcolor{black}{Therefore, a single SP with multiple slice tenants is adopted in this research, the overall architecture is shown in Fig.~\ref{fig:architecture}.}

\textcolor{black}{Heterogeneous requests are proposed by subscribers and sent to the corresponding tenants that can serve them. If there are too many requests backed up in the queue, new requests may balk from their slice tenants. Those successfully submitted to the tenants will stay in queues to wait for service until their proposers run out of patience and withdraw them. Resources are allocated to slices according to the QoS requirements of ongoing requests and eventually taken back by the SP when requests are finished.}

Slice tenants are internet service providers who have no substrate devices, and sell virtual network businesses to their subscribers by leasing tailored virtual slices from the SP. One SP can deliver slices to several tenants at the same time with a pre-negotiated service level agreement (SLA), earn from it, and pay a penalty if the SLA is violated. Assuming that there are $K$ tenants signed a contact with the SP, and are included in a set $\mathcal{K} = \left\{ 1, 2, \dots, K \right\}$. Each tenant corresponds to one type of service, which consumes multiple kinds of resources, such as CPU, GPU, bandwidth, memory, storage, et al. Supposing there are total $N$ types of resource needed by all tenants under the SP, then the resource demand of tenant $k$ can be formulated with a $N$ dimensional non-negative column vector $\boldsymbol{r}_k = \left( r^1_k, r^2_k, \dots, r^N_k \right)^T$, where each element denotes the amount of each resource. For elastic services, the demand vector gives the average resource demands during their lifetime, while for inelastic services, this marks the real-time demands that should be held all the time until expiration. We use a $N \times K$ matrix $\boldsymbol{R} = \left\{ \boldsymbol{r}_1, \boldsymbol{r}_2, \dots, \boldsymbol{r}_K \right\}$ to denote the resource demands of all the tenants, and the total resource capacities possessed by the SP is $\boldsymbol{c} = \left( c^1, c^2, \dots, c^N \right)^T$.

Requests can obtain a priority level that aligns with that of the slice serving them. However, there remains a puzzle. For a part of tenants, requests inside ought to have different priorities due to the mobility of service users, i.e., handoff requests should be prior to new requests for the same service type. To solve this problem, a small supplementary is required. We treat tenants that may receive handoff requests as two sub-tenants with the same demand vector but different urgencies, one for new requests and another for handoff requests. The same adjustments are made to the corresponding slices and a higher priority is appointed to the more urgent slice. We slightly abuse the set $\mathcal{K} = \left\{ 1, 2, \dots, K \right\}$ to label the priority level of each tenant and sub-tenant or each slice and sub-slice equivalently. The bigger the number, the higher the priority level is.

The SP establishes a queue for each slice to accommodate requests proposed by tenants that cannot be served immediately for a temporary shortage of resources. Requests inside a queue have the same demand vector and priority level, so the First Come First Serve (FIFO) principle is employed within each queue when resources are sufficient to pull requests from it. Time is divided into slots that are fine enough to work as a basic unit to measure the lifetime and hold time of requests. For a request $i$, we use $T^{\rm L}_i$ and $T^{\rm H}_i$ to record its lifetime and holdtime durations respectively. These two parameters can even differ between requests in the same queue, and only $T^{\rm L}$ of elastic requests is assumed revealed to the SP at the moment of initiation by their proposers, the others are invisible. We introduce another symbol $T^{\rm P}_i$ to indicate the perceived time of request $i$ by the SP, i.e., the time slot at which this request joins the corresponding queue. Then, the renege behavior of subscribers can be characterized when the Ineq.~\eqref{eq:renege} is not held, where $t$ indicates the current time slot sequence. Vector $\boldsymbol{d}(t) = \left( d_1(t), d_2(t), \dots, d_K(t) \right)^T$ is used to denote the number of requests depart from each queue because of renege.

\begin{equation} \label{eq:renege}
	T^{\rm P}_i + T^{\rm H}_i - 1 \leq t
\end{equation}

We use $\mathcal{O}(t)=\underset{k \in \mathcal{K}}{\cup}\mathcal{O}_k(t)$ to represent the set of all ongoing requests at time slot $t$, and $\mathcal{O}_k(t)$ corresponds to those in slice $k$. While, $\boldsymbol{l}(t) = \left( l_1(t), l_2(t), \dots, l_K(t) \right)^T$ is a vector that records the length of all slices at slot $t$, each element matches a specific queue, reflecting the number of requests waiting for service. The admission decision must take into account the status of queues, i.e., $\boldsymbol{l}(t)$, and available resources, i.e., $\boldsymbol{c}^{\rm ava}(t)$. The symbol $\boldsymbol{r}$ is multiplexed to define a function $\boldsymbol{r}(\cdot)$ that takes a request $i$ as input and returns its demand vector if it has not yet been served, or its resources actually allocated when it is an ongoing request. Thus, $\boldsymbol{c}^{\rm ava}(t)$ can be computed through Eq.~\eqref{eq:c^ava}.

\begin{equation} \label{eq:c^ava}
	\boldsymbol{c}^{\rm ava}(t) = \boldsymbol{c} - \sum_{i \in \mathcal{O}(t)}{\boldsymbol{r}(i)}
\end{equation}

At the beginning of each slot, some ongoing requests may saturate their lifetime, expire from the system and release the resources they previously occupied. We formulate the condition for these operations briefly as Ineq.~\eqref{eq:expire}, where $T^{\rm A}_i$ is the slot at which request $i$ is admitted by the SP. Note that $T^{\rm A}_i$ must satisfy the Inequality~\eqref{eq:condition for T^A_i}, because the renege behavior is considered. Similar to $\mathcal{O}(t)$, we use $\mathcal{E}(t)=\underset{k \in \mathcal{K}}{\cup}\mathcal{E}_k(t)$ to indicate the set of those requests died out at time slot $t$. Then, the ongoing set $\mathcal{O}(t)$ and available resources $\boldsymbol{c}^{\rm ava}(t)$ should be updated according to Eqs.~\eqref{eq:update ongoing set} and~\eqref{eq:update c^ava} respectively. Besides, we stimulate the balk behavior of subscribers by attaching a probability to each slice $k$ that decreases with the increase of its queue length, see Eq.~\eqref{eq:balk probability}, $\beta_k \in [0,1]$ is a slice dependent parameter to show the willingness of subscribers to wait for type-$k$ slice~\cite{han2019utility}. New requests initiated by users will join the queue with probability $p_k(t)$ rather than entering directly with no hesitation. It's easy to check that when there are no requests waiting in the queue, new ones will definitely join. But if the queue length is too long, they rarely join, because this usually means an unacceptable waiting time that elevates the risk of violation in Ineq.~\eqref{eq:renege}, i.e., renege. We use another vector $\boldsymbol{b}(t) = \left( b_1(t), b_2(t), \dots, b_K(t) \right)^T$ to count the number of requests lost at time slot $t$, due to the balk behavior.

\begin{equation} \label{eq:expire}
	T^{\rm A}_i + T^{\rm L}_i \leq t
\end{equation}
\begin{equation} \label{eq:condition for T^A_i}
	T^{\rm P}_i \leq T^{\rm A}_i \leq T^{\rm P}_i + T^{\rm H}_i - 1
\end{equation}
\begin{equation} \label{eq:update ongoing set}
	\mathcal{O}(t) = \mathcal{O}(t) \backslash \mathcal{E}(t)
\end{equation}
\begin{equation} \label{eq:update c^ava}
	\boldsymbol{c}^{\rm ava}(t) = \boldsymbol{c}^{\rm ava}(t) + \sum_{i \in \mathcal{E}(t)}{\boldsymbol{r}(i)}
\end{equation}
\begin{equation} \label{eq:balk probability}
	p_k(t) = e^{-\beta_k \cdot l_k(t)} \quad ,\forall k \in \mathcal{K}
\end{equation}

The expiration of ongoing requests is captured by the SP as soon as they set resources free, while the loss of requests because of reneging and balking is reported by tenants to the SP periodically, i.e., each time slot. With this information, the SAC algorithm is responsible for deciding which kind of requests and how many of them should be newly admitted at that exact time slot. We use a vector $\boldsymbol{a}(t) = \left( a_1(t), a_2(t), \dots, a_K(t) \right)^T$ to term the admission actions, $a_k(t)$ means that the top $a_k(t)$ requests in type-$k$ queue are to be pulled to service.

%\begin{figure}[!t]
%	\centering
%	\includegraphics[scale=0.5]{architecture}
%	\caption{System architecture}
%	\label{fig:architecture}
%\end{figure}

\subsection{Formulation of priority requirement}

In our design, the priority levels among slices are reflected through the measurement of the cumulative service acceptance ratio (CSAR) of each kind of service. It is defined as the quotient of the cumulative number of admitted requests, termed as $cum^{\rm A}_{k}(t)$, and the cumulative number of confirmed requests, termed as $cum^{\rm C}_{k}(t)$. The latter refers to those who obtained a definite admission result, including admitted, reneged, and balked. Eq.~\eqref{eq:cum^A} and Eq.~\eqref{eq:cum^C} give the corresponding formulations of them. Then, the CSAR of type-$k$ slice at time slot $t$, termed as $\alpha_k(t)$, can be computed by Eq.~\eqref{eq:CSAR}. We apply this to each slice and form a CSAR vector $\boldsymbol{\alpha}(t)=\left(\alpha_1(t), \alpha_2(t), \dots, \alpha_K(t) \right)^T$.

\begin{equation} \label{eq:cum^A}
	cum^{\rm A}_{k}(t) = \sum_{\tau = 1}^{t}{a_{k}(\tau)} \quad ,\forall k \in \mathcal{K}
\end{equation}
\begin{equation} \label{eq:cum^C}
	cum^{\rm C}_{k}(t) = \sum_{\tau = 1}^{t}{a_{k}(\tau) + d_{k}(\tau)+ b_{k}(\tau)} \quad ,\forall k \in \mathcal{K}
\end{equation}
\begin{equation} \label{eq:CSAR}
	\alpha_k(t) = \frac{cum^{\rm A}_{k}(t)}{cum^{\rm C}_{k}(t)} \quad ,\forall k \in \mathcal{K}
\end{equation}

Without loss of generality, we assume the priority level of each slice is labeled by its index, so the priority requirement is met if the vector $\boldsymbol{\alpha}(t)$ is monotonous non-decrease, i.e., sequential Inequalities~\eqref{eq:priority reqiurement} holds. We introduce an indicator function $\mathcal{I}(\cdot)$ that takes $\boldsymbol{\alpha}(t)$ as input, outputs $1$ if $\boldsymbol{\alpha}(t)$ satisfies the priority requirement, $0$ otherwise. Eq.~\eqref{eq:indicator function} gives a mathematical description of $\mathcal{I}(\cdot)$.

\begin{equation} \label{eq:priority reqiurement}
	\alpha_1(t) \leq \alpha_2(t) \leq \cdots \leq \alpha_K(t)
\end{equation}
\begin{equation} \label{eq:indicator function}
	\mathcal{I}(\boldsymbol{\alpha}(t)) = 
	\left\{ 
	\begin{array}{ll}
		1, & \text{if } \alpha_{k}(t) \leq \alpha_{k+1}(t), \forall k \in [1,K) \\
		
		0, & \text{otherwise}
	\end{array} 
	\right.
\end{equation}

The above formulations just give a rough picture about whether the priority is preserved. To evaluate the degree of priority between adjacent levels, gap of CSARs is a good choice, we term it as a column vector $\Delta \boldsymbol{\alpha}(t) = \left( \Delta \alpha_1(t), \Delta \alpha_2(t), \dots, \Delta \alpha_{K-1}(t) \right)^T$ with $K$-1 dimensions, where $\Delta \alpha_k(t) \triangleq \alpha_{k+1}(t) - \alpha_k(t)$ matches the gap between slice $k$ and slice $k$+1.

Another confusing problem that needs further explanation is when the priority condition is upset, which slice is indeed responsible for this. We elaborate this with two CSAR vectors $\boldsymbol{\alpha}(t_1) = (0.7, 0.8, 0.9, 0.85)^T$ and $\boldsymbol{\alpha}(t_2) = (0.95, 0.8, 0.9, 0.92)^T$. In slot $t_1$, it's the fourth slice that breaks the requirement, if we can lift it to 0.9 or even higher, Inequality~\eqref{eq:priority reqiurement} holds. Then, does the first slice is the one to blame in slot $t_2$ for such a high CSAR, the answer is no. Actually, in $\boldsymbol{\alpha}(t_2)$, the breakers are slice 2, slice 3, and slice 4. If we treat slice 1 as the culprit, depression in its CSAR is needed to fix the whole vector, it means some ongoing requests of its type must be dropped immediately to decrease the numerator in Eq.~\eqref{eq:CSAR}, which is a taboo in SLA-guaranteed scenario, and is neither allowed in our SAC algorithm. Therefore, slices that violate the priority requirement refer to those whose CSARs can be promoted to turn previously unsatisfied Ineq.~\eqref{eq:priority reqiurement} into satisfactions.

\subsection{Formulation of fairness degree}

In a prioritized environment, giving more preference on high-priority slices is not only more reasonable but also more desirable. However, the degree of this preference should not be stretched indefinitely and deserves to be limited within a certain range, this is why and where we impose the fairness objective.

The relative degrees of priority between adjacent levels are not necessarily consistent with each other, a weight vector $\boldsymbol{w}=\left( w_1, w_2, \dots, w_{K-1} \right)^T$ is used as a parameter to coordinate the ideal relative size of them, each matches an element in $\Delta \boldsymbol{\alpha}(t)$. It can be tailored according to the actual slices offered by the SP. When the tenant population changes, the SP can flexibly adapt to it by adding, removing, or adjusting the corresponding weight values.

{\color{black}Therefore, we hope the new definition of fairness can be sensitive to whether the priority requirement is met and reflect the correlation between $\Delta \boldsymbol{\alpha}(t)$ and $\boldsymbol{w}$. The uniformity degree of the zoomed CSAR gaps, termed $f_{J}(\boldsymbol{\alpha}(t))$, is exactly the one to make this. We make minor adjustments to \textit{Jain}'s fairness index~\cite{jain1984quantitative} and quantity it in Eq.~\eqref{eq:fairness}}.

%The peak value of fairness is obtained when vector $\Delta \boldsymbol{\alpha}(t)$ is totally proportional to vector $\boldsymbol{w}$. On the contrary, the violation of the priority requirement is considered to be the most unfair thing. 

%Therefore, the degree of uniformity of the zoomed CSAR gaps can directly reflect the degree of fairness. This statement conveys the definition of fairness rigorously, but is not intuitive enough, a quantization method is essential for us to judge (with easy effort) which admission result is fairer. We make minor adjustments to \textit{Jain}'s fairness index~\cite{jain1984quantitative} to let it fit into our scenario. Details are shown in Eq.~\eqref{eq:fairness}.

\begin{equation} \label{eq:fairness}
	f_{J}(\boldsymbol{\alpha}(t)) =
	\left\{ 
	\begin{array}{cl}
		\frac{ \left( \sum_{k=1}^{K-1}{ \frac{\Delta \alpha_k(t)}{w_k} } \right)^2 }
		{ \left( K-1 \right) \cdot \sum_{k=1}^{K-1}{ \left( \frac{\Delta \alpha_k(t)}{w_k} \right)^2} }, & \text{if } \mathcal{I} \left( \boldsymbol{\alpha}(t) \right) \\
		0, & \text{otherwise}
	\end{array} 
	\right.
\end{equation}

{\color{black}
It can be checked that function $f_{J}(\boldsymbol{\alpha}(t))$ is non-negative and limited to $\left[0,1\right]$. A violation of the priority requirement will suppress $f_{J}(\boldsymbol{\alpha}(t))$ to zero. Meanwhile, other positive values can directly reveal the degree of fairness, which is an important property of \textit{Jain}'s fairness index and inherited by $f_{J}(\boldsymbol{\alpha}(t))$.
% What's more, the following theorem gives the condition that makes $f_{J}(\boldsymbol{\alpha}(t))$ reach its peak value.

%\newtheorem{thm}{\bf Theorem}
%\begin{thm}\label{thm:peak fairness condition}
%	The fairness indicator takes its upper bound if and only if the CSAR gap vector $\Delta \boldsymbol{\alpha}(t)$ is proportional to the weight vector $\boldsymbol w$.
%\end{thm}
%
%\begin{equation} \label{eq:jain index}
%	f(\boldsymbol{x})=\frac{\left[ \sum_{i=1}^{n}{x_i} \right]^2}{n \cdot \sum_{i=1}^n{x_i^2}}, \quad x_i \ge 0
%\end{equation}
%
%\begin{equation} \label{eq:peak fairness condition}
%	\frac{\Delta \alpha_i(t)}{w_i} = \frac{\Delta \alpha_j(t)}{w_j}, \quad \forall i \neq j \in \left[1,K\right)
%\end{equation}
%
%\begin{proof}
%	It has been proved in \cite{jain1984quantitative} that the {\it Jain}'s fairness index, shown in Eq.~\eqref{eq:jain index}, obtains the maximum value 1 only when $x_i$ equals to each other. Substitute $x_i$ with $\Delta \alpha_k(t)/w_k$, we can get that $f_{J}(\boldsymbol{\alpha}(t))$ reaches 1 if and only if equations in Eq.~\eqref{eq:peak fairness condition} hold. Through a simple transformation, it can be expressed as $\Delta \boldsymbol{\alpha}(t) = \epsilon \cdot \boldsymbol{w}$, where $\epsilon$ is the proportion factor. Q.E.D.
%\end{proof}

}

%It can be checked that function $f_{J}(\boldsymbol{\alpha}(t))$ is non-negative, and obtains its maximum value $1$ when the vector $\boldsymbol{\alpha}(t)$ is strictly proportional to vector $\boldsymbol{w}$, reaches the bottom 0 when priority is violated, regardless of the dimension and absolute value of the input. In other words, the output of $f_{J}(\cdot)$ always lies within the range $[0,1]$ under any input samples, and the value taken is positively correlated with the degree of fairness, i.e., the properties of original \textit{Jain}'s fairness index is inherited by $f_{J}(\cdot)$. 

\subsection{Optimization problem}

The SAC module needs three parts of state information to make an admission decision, the set of ongoing requests $\mathcal{O}(t)$, the vector of real-time queue length $\boldsymbol{l}(t)$ and the initial CSAR vector $\boldsymbol{\alpha}^{\rm I}(t)$. The first two are used to determine the available resource capacities and the total number of requests in queues at time slot $t$, respectively. The third part is computed as Eq.~\eqref{eq:Initial CSAR}, where vector division is defined as the division of counterpart elements. It stands for the initial CSAR of each slice at the beginning of slot $t$, where admission decisions have not been made and executed yet, but the behavior of reneging and balking had already happened. 

\begin{equation} \label{eq:Initial CSAR}
	\boldsymbol{\alpha}^{\rm I}(t) = \frac{\boldsymbol{cum}^{\rm A}(t-1)}{\boldsymbol{d}(t)+\boldsymbol{b}(t)+\boldsymbol{cum}^{\rm C}(t-1)}
\end{equation}
\begin{equation} \label{eq:policy}
	\boldsymbol{\pi}:\left\{ \mathcal{O} \right\} \times \left\{ \boldsymbol{l} \right\} \times \left\{ \boldsymbol{\alpha}^{\rm I} \right\} \mapsto \mathbb{R}^K_+
\end{equation}
\begin{equation} \label{eq:utilization}
	u(t) = {\rm Min} \, \left\{ \frac{\sum_{i \in \mathcal{O}(t)}{\boldsymbol{r}(i)}}{\boldsymbol{c}} \right\}
\end{equation}

In mathematical terms, slice admission control is an operation, denoted as $\boldsymbol{\pi}$, for implementing a mapping from the state tuple to an action vector, illustrated in Eq.~\eqref{eq:policy}. During the process, resource constraints should be maintained and some objectives are pursued, i.e., mainly the priority requirement and fairness in this work. Utilization of resources, see Eq.~\eqref{eq:utilization}, is also taken into account to avoid too many idle resources. The final model of our optimization problem is shown in Eq.~\eqref{eq:objectives}.

\begin{equation} \label{eq:objectives}
	\underset{\boldsymbol{a}(t)}{\text{Maximize}} \quad \left(
	\frac{\sum_{\tau = 1}^{t}{\mathcal{I}\left( \boldsymbol{\alpha}(\tau) \right)}}{t},
	f_{J}\left( \boldsymbol{\alpha}(t) \right),
	u(t)
	\right)
\end{equation}
\begin{align}
	s.t. \quad 
	\boldsymbol{a}(t)  & \gets \boldsymbol{\pi}\left( \mathcal{O}(t), \boldsymbol{l}(t), \boldsymbol{\alpha}^{\rm I}(t) \right) \tag{\ref{eq:objectives}a} \label{eq:constraint_a}\\
	\boldsymbol{a}(t) & \leq \boldsymbol{l}(t) \tag{\ref{eq:objectives}b} \label{eq:constraint_b} \\
	\boldsymbol{R} \cdot \boldsymbol{a}(t) & \leq \boldsymbol{c}^{\rm ava}(t) \tag{\ref{eq:objectives}c}  \label{eq:constraint_c}	
\end{align}

The first objective is the time average of the sum of historical priority requirements, which means we aim for a long-term smooth priority performance. Other objectives have received plenty of explanations before, we do not bother anymore.  Constraint~\eqref{eq:constraint_a} indicates that the admission action originates from the current system state, constraint~\eqref{eq:constraint_b} represents that the number of requests to access cannot exceed the amount accumulated in queues, constraint~\eqref{eq:constraint_c} implies that the total resource demands of these newly accepted requests should be affordable to the residual resources. Symbol $t$ indicates that this is an online problem with different system states and constraints at different time slots.

%% file: section/sec05.AlgorithmDesign.tex
\section{Algorithm Design} \label{algorithm design}

\textcolor{black}{The problem formulated in Eq.~\eqref{eq:objectives} is obviously non-convex, and the online character makes time-varying constraints, which further complicates the issue. Thus, it can hardly be solved with the classical mathematical optimization methods in a time-efficient way. Reinforcement learning methods are also not suitable for solving this problem, as multiple optimization objectives may lead to unstable model convergence. What's worse, adjustments to tenant types and alterations in the resource demands of tenants require reconstruction of the neural network structure and retraining of the neural network parameters, which hinders a timely and accurate decision output sometimes. Therefore, we propose a heuristic-based SAC algorithm called PSACCF, it introduces the resource efficiency of services to amend priority violations, then promotes fairness by setting the target CSARs for each service type and pushing the admission result to approach the optimal solution.}

%\begin{algorithm}[!th] 
%	\caption{PSACCF} \label{alg:PSACCF}
%	\SetKwInOut{Input}{Input}
%	\SetKwInOut{Output}{Output}
%	
%	\Input{1) Set of ongoing requests $\mathcal{O}(t)$; 2) Length of queues $\boldsymbol{l}(t)$; 3) Initial CSAR vector $\boldsymbol{\alpha}^{\rm I}(t)$.}
%	\Output{Admission decision $\boldsymbol{a}(t)$.}
%	Initiate $\boldsymbol{a}(t)=0$, $\boldsymbol{\alpha}(t)=\boldsymbol{\alpha}^{\rm I}(t)$, count $\boldsymbol{d}(t)$ and $\boldsymbol{b}(t)$\;
%	Update $\boldsymbol{l}(t)$, $\boldsymbol{c}^{\rm ava}(t)$ and $\boldsymbol{DR}(t)$\;
%	\tcp{priority amendment}
%	Execute \textbf{Procedure~\ref{alg:fix priority}} \;
%	\If{$\mathcal{I}(\boldsymbol{\alpha}(t))==1$}{
%		\tcp{fairness enhancement}
%		\While{true}{
%			Execute \textbf{Procedure~\ref{alg:check fairnes}} \tcc*{check the degree of fairness}
%			Execute \textbf{Procedure~\ref{alg:calculate target CSAR}} \tcc*{calculate target CSAR}
%			Execute \textbf{Procedure~\ref{alg:approach to target CSAR}} \tcc*{approach to target CSAR}
%			\If{$\boldsymbol{\alpha}(t)$ remains unchanged after {\rm \textbf{Procedure~\ref{alg:approach to target CSAR}}}}{
%				\textbf{break}\;
%			}
%		}
%	}
%	\textbf{return} $\boldsymbol{a}(t)$\;
%\end{algorithm}

\subsection{Admission decision}

{\color{black}It can be inferred from Eq.~\eqref{eq:fairness} that the degree of fairness relies on whether the priority requirement is satisfied, so PSACCF can roughly be divided into two parts, priority amendment and fairness enhancement.}

\subsubsection{Priority amendment} \ \par

{\color{black}
PSACCF starts with an inspection of the priority requirement, queues violating it are added into the set $\mathcal{V}(t)$. Requests belonging to the queues within $\mathcal{V}(t)$ need to be processed first to correct the priority. The resource efficiency, denoted as $RE_k(t)$, is introduced to determine the admission order of these requests.

\begin{equation} \label{eq:resource efficiency}
	RE_k(t) = \frac{\partial \alpha_k(t)}{\partial a_k(t)} / \boldsymbol{r}^{DR_k(t)}_k %= \frac{1-\alpha_k(t)}{\left( 1+cum^{\rm C}_k(t) \right) \cdot \boldsymbol{r}^{DR_k(t)}_k}
\end{equation}

$RE_k(t)$ is formulated in Eq.~\eqref{eq:resource efficiency}, where $\partial \alpha_k(t) / \partial a_k(t)$ denotes the derivative of CSAR for each queue with respect to its access volume. It measures the incremental increase in the CSAR of the service for each new incoming request, shown in Eq.~\eqref{eq:partial alpha}. The denominator is the amount of dominant resource acquired by the request, whose superscript $DR_k(t)$ indicates the dominant label and is judged by Eq.~\eqref{eq:DR}. It marks the label of the first saturated resource when $\boldsymbol{c}^{\rm idle}(t)$ in Eq.~\eqref{eq:c^idle} is used to accommodate type-$k$ requests exclusively.
}
%To determine the admission order, PSACCF first examines the derivative of CSAR for each queue with respect to its access volume, shown in Eq.~\eqref{eq:partial alpha}. {\color{black}It measures the incremental increase in the CSAR of the service for each new incoming request, so the larger the value is, the higher the amount of priority correction will be.}
\begin{equation} \label{eq:partial alpha}
	\frac{\partial \alpha_k(t)}{\partial a_k(t)} = \frac{1+cum^{\rm A}_k(t)}{1+cum^{\rm C}_k(t)} - \alpha_k(t) = \frac{1-\alpha_k(t)}{1+cum^{\rm C}_k(t)}
\end{equation}

\begin{equation} \label{eq:DR}
	DR_k(t) = {\rm argmax}_n \left(
	\frac{\boldsymbol{r}_k}{\boldsymbol{c}^{\rm idle}(t)}
	\right)
\end{equation}
\begin{equation} \label{eq:c^idle}
	\boldsymbol{c}^{\rm idle}(t) = \boldsymbol{c}^{\rm ava}(t) - \boldsymbol{R} \cdot \boldsymbol{a}(t)
\end{equation}

{\color{black}
The resource efficiency reflects the size of the increment in CSAR resulting from the consumption of the same amount of dominant resource. Therefore, the higher $RE_k(t)$ is, the faster the priority amendment it brings, which can act as a criterion to sort requests.

\begin{algorithm}
	\renewcommand{\algorithmcfname}{Procedure}
	\renewcommand{\thealgocf}{1}
	\caption{Fix the priority requirement} \label{alg:p1 fix priority}
	\While{$\mathcal{I}(\boldsymbol{\alpha}(t))==0$ \label{alg:p1 priority amendment begin}}{
		Construct $\mathcal{V}(t)$, descending sort $\mathcal{V}(t)$ with $\boldsymbol{RE}(t)$\; \label{alg:p1 construct violation set}
		\For{each $k \in \mathcal{V}(t)$ }{
			\If{$\boldsymbol{R} \cdot (\boldsymbol{a}(t) + \boldsymbol{e}_k) \leq \boldsymbol{c}^{\rm ava}(t)$}{
				$a_k(t)$++, $l_k(t)${-}{-}, update $\alpha_k(t)$\;
				\textbf{break}\;
			}
		}
		\If{$\boldsymbol{\alpha}(t)$ remains unchanged above}{
			\textbf{break}\; \label{alg:p1 priority amendment end}
		}
	}
\end{algorithm}

{\bf Procedure~\ref{alg:p1 fix priority}} shows the process of priority amendment. Each time when the priority requirement is unsatisfied, the set $\mathcal{V}(t)$ is formed, and members inside are arranged in descending order for their resource efficiencies. At which sequence the resource availability is checked to see whether there exists a queue $k$ that admitting one request will not violate the resource constraint. If so, the first request in this queue will be pre-admitted for CSAR update. These steps repeat until there are no such queues or the priority requirement is satisfied.
}

\subsubsection{Fairness enhancement} \ \par

{\color{black}
Once the priority requirement is satisfied, PSACCF enters the phase of fairness enhancement. A threshold $\varphi \in [0,1]$ is used to control the acceptable degree of fairness. When $f_{J}(\boldsymbol{\alpha}(t))$ exceeds $\varphi$, it is regarded as fair enough so that admission actions can be skewed towards requests with high priority to further consolidate the priority requirement. That is exactly the job of {\bf Procedure~\ref{alg:check fairnes}}. Line~\ref{alg:p2 for begin} to Line~\ref{alg:p2 for end} try to admit requests in a high priority first manner and are repeated until there are no such requests or the fairness indicator downs to an inadequate level.
}

%After the process of priority amendment, efforts are going to be made to promote the degree of fairness. Before this, the priority requirement is twice inspected, if the result is still negative, then it means that the available resources are not sufficient to serve extra requests, indicating that optimization for fairness makes no sense. If the priority requirement is achieved when priority amendment is done, fairness enhancement begins.

\begin{algorithm}
	\renewcommand{\algorithmcfname}{Procedure}
	\renewcommand{\thealgocf}{2}
	\caption{Check the degree of fairness} \label{alg:check fairnes}
	\While{$f_{J}(\boldsymbol{\alpha}(t)) \geq \varphi$  \label{alg:fair enough begin}}{
		\For{each $k \in \left\{ K,\dots,1 | l_k(t)>0 \right\}$ \label{alg:p2 for begin}}{
			\If{$\boldsymbol{R} \cdot (\boldsymbol{a}(t) + \boldsymbol{e}_k) \leq \boldsymbol{c}^{\rm ava}(t)$}{
				$a_k(t)$++, $l_k(t)${-}{-}, update $\alpha_k(t)$\;
				\textbf{break}\;
			}
		} \label{alg:p2 for end}
		\If{$\boldsymbol{\alpha}(t)$ remains unchanged above}{
			\textbf{break}\;  \label{alg:fair enough end}
		}
	}
\end{algorithm}

{\color{black}
When the degree of fairness is not high enough, efforts are going to be made to improve it. First of all, the target CSARs (termed as $\boldsymbol{\alpha}^{\rm tar}(t)$) of each kind of service, under which the fairness indicator reaches 1, should be derived in advance. The following theorem provides the fundamental basis for calculating the target CSARs.

\newtheorem{thm}{\bf Theorem}
\begin{thm}\label{thm:peak fairness condition}
	The fairness indicator takes its upper bound if and only if the CSAR gap vector $\Delta \boldsymbol{\alpha}(t)$ is proportional to the weight vector $\boldsymbol w$.
\end{thm}

\begin{equation} \label{eq:jain index}
	f(\boldsymbol{x})=\frac{\left[ \sum_{i=1}^{n}{x_i} \right]^2}{n \cdot \sum_{i=1}^n{x_i^2}}, \quad x_i \ge 0
\end{equation}

\begin{equation} \label{eq:peak fairness condition}
	\frac{\Delta \alpha_i(t)}{w_i} = \frac{\Delta \alpha_j(t)}{w_j}, \quad \forall i \neq j \in \left[1,K\right)
\end{equation}

\begin{proof}
	It has been proved in \cite{jain1984quantitative} that the {\it Jain}'s fairness index, shown in Eq.~\eqref{eq:jain index}, obtains the maximum value 1 only when $x_i$ equals to each other. Substitute $x_i$ with $\Delta \alpha_k(t)/w_k$, we can get that $f_{J}(\boldsymbol{\alpha}(t))$ reaches 1 if and only if equations in Eq.~\eqref{eq:peak fairness condition} hold. Through a simple transformation, it can be expressed as $\Delta \boldsymbol{\alpha}(t) = \epsilon \cdot \boldsymbol{w}$, where $\epsilon$ is the proportion factor. Q.E.D.
\end{proof}

}

%PSACCF sets a threshold $\varphi \in [0,1]$ to denote the acceptable degree of fairness. It is regarded as fair enough when the result of Eq.~\eqref{eq:fairness} exceeds this threshold. During this stage, admission actions can be skewed towards high priority services to further consolidate the priority requirement, see 
%%Lines~\ref{alg:fair enough begin}-\ref{alg:fair enough end} in Alg.~\ref{alg:PSACCF}. 
%{\bf Procedure~\ref{alg:check fairnes}}. If the degree of fairness is not as high as desired, optimization follows closely behind.

%In our design, the idea for enhancing fairness comes from the characteristic of the original \textit{Jain}'s fairness index, which gets its maximum value when each participant equals to the others. Putting it into our model, $f_{J}(\cdot)$ reaches the peak value when Eq.~\eqref{eq:peak fairness condition} holds. This principle inspires us to set a target CSAR value, termed as $\alpha_k^{\rm tar}(t)$, for each queue. Another parameter $\epsilon$ is employed to control the actual gap of CSAR between adjacent queues, which allows PSACCF to optimize fairness without sacrifice the priority differences.

\begin{algorithm}
	\renewcommand{\algorithmcfname}{Procedure}
	\renewcommand{\thealgocf}{3}
	\caption{Calculate the target CSAR} \label{alg:p3 calculate the tar CSAR}
	\eIf{$\alpha_K(t) - \alpha_1(t) > \epsilon \cdot \sum_{i=1}^{K-1}{w_i}$ \label{alg:p3 check CSAR span}}{
		\For{each $k \in \mathcal{K}$}{
			$\alpha^{\rm tar}_k(t)=\alpha_K(t) - \epsilon \cdot \sum_{i=k}^{K-1}{w_i}$\;
		} \label{alg:p3 tar CSAR v1 end}
	}{
		\eIf{$\alpha_1(t)+\epsilon \cdot \sum_{i=1}^{K-1}{w_i} \leq 1$ \label{alg:p3 check tar CSAR upper bound}}{
			\For{each $k \in \mathcal{K}$}{
				$\alpha^{\rm tar}_k(t) = \alpha_1(t) + \epsilon \cdot \sum_{i=1}^{k-1}{w_i}$\;
			} \label{alg:p3 tar CSAR v2 end}
		}{ \label{alg:p3 tar CSAR upper bound exceed}
			\For{each $k \in \mathcal{K}$}{
				$\alpha^{\rm tar}_k(t) = {\rm max} \left\{ 1 - \epsilon \cdot \sum_{i=k}^{K-1}{w_i}, 0 \right\}$\; \label{alg:p3 tar CSAR v3 end}
			}
		}
	}
\end{algorithm}

{\color{black}
Based on {\bf Theorem~\ref{thm:peak fairness condition}}, the principle for calculating the target CSARs is illustrated in {\bf Procedure~\ref{alg:p3 calculate the tar CSAR}}. The distance between the actual CSARs of the least prioritized and the most prioritized service is compared with the weighted sum of $\epsilon$. A greater distance means the span of $\Delta \boldsymbol{\alpha}(t)$ can be further narrowed by promoting the target CSARs of those less preferred services, which can be realized using Eq.~\eqref{eq:tar alpha v1} and shown in Line~\ref{alg:p3 check CSAR span} to Line~\ref{alg:p3 tar CSAR v1 end}. Otherwise, the target CSARs of those more prioritized services can be raised to extend the distance between $\alpha_1(t)$ and $\alpha_K(t)$. Since the CSAR is bounded between 0 and 1, a secondary judgment is executed in Line~\ref{alg:p3 check tar CSAR upper bound}. Different update formulas are adopted accordingly, which are listed in Eq.~\eqref{eq:tar alpha v2} and Eq.~\eqref{eq:tar alpha v3}, pseudocoded in Line~\ref{alg:p3 check tar CSAR upper bound} to Line~\ref{alg:p3 tar CSAR v2 end} and Line~\ref{alg:p3 tar CSAR upper bound exceed} to Line~\ref{alg:p3 tar CSAR v3 end} respectively. During this procedure, the proportion factor $\epsilon$ works as a vital parameter to control the absolute CSAR gaps between all adjacent priority levels.
}

%To get the target CSAR of each queue, the span of actual CSAR between the least prioritized queue and the most prioritized queue is compared with the weighted sum of $\epsilon$, 
%%i.e., Lines~\ref{alg:span judge} in Alg.~\ref{alg:PSACCF}, 
%where $\epsilon$ is a parameter between $[0,1]$ to control the  gap of CSARs between adjacent priority levels. If the span is bigger, it means $\boldsymbol{\alpha}(t)$ of all these services can be further narrowed, which can be realized by promoting the CSAR of those less preferred queues. Thus, $\alpha_k^{\rm tar}(t)$ of each service can be computed as Eq.~\eqref{eq:update alpha v1} in this circumstance. Otherwise, the span should be extended by raising the CSAR of those preferred queues. While, as CSAR is bounded between 0 and 1, formula for calculating $\boldsymbol{\alpha}^{\rm tar}(t)$ should be separated into Eq.~\eqref{eq:update alpha v2} and Eq.~\eqref{eq:update alpha v3} in the latter case, based on whether the sum of $\alpha_1(t)$ and the weighted sum of $\epsilon$ is limited to 1. The process of updating $\boldsymbol{\alpha}^{\rm tar}(t)$ is shown in 
%%Lines~\ref{alg: target alpha begin}-\ref{alg: target alpha end} of Alg.~\ref{alg:PSACCF}
%{\bf Procedure~\ref{alg:calculate target CSAR}}.
\begin{equation} \label{eq:tar alpha v1}
	\alpha^{\rm tar}_k(t)=\alpha_K(t) - \epsilon \cdot \sum_{i=k}^{K-1}{w_i}, \quad \forall k \in \mathcal{K}
\end{equation}
\begin{equation} \label{eq:tar alpha v2}
	\alpha^{\rm tar}_k(t) = \alpha_1(t) + \epsilon \cdot \sum_{i=1}^{k-1}{w_i}, \quad \forall k \in \mathcal{K}
\end{equation}
\begin{equation} \label{eq:tar alpha v3}
	\alpha^{\rm tar}_k(t) = {\rm max} \left\{ 1 - \epsilon \cdot \sum_{i=k}^{K-1}{w_i}, 0 \right\}, \quad \forall k \in \mathcal{K}
\end{equation}

{\color{black}
$\boldsymbol{\alpha}^{\rm tar}(t)$ generated by {\bf Procedure~\ref{alg:p3 calculate the tar CSAR}} exactly satisfies the condition illustrated in {\bf Theorem \ref{thm:peak fairness condition}}, so it provides an ideal reference value for each actual CSAR to pursue. Admission decisions should push $\boldsymbol{\alpha}(t)$ toward $\boldsymbol{\alpha}^{\rm tar}(t)$ so that the fairness can be optimized. Considering that dropping an ongoing request will cause a penalty in the SlaaS environment, only those whose actual CSAR is lower than the target CSAR can become the ones for admission. Another set $\mathcal{P}(t)$ is used to record this kind of service.
}

%$\boldsymbol{\alpha}^{\rm tar}(t)$ offers a state for each service, under which the fairness of the system reaches the optimum and the priority differences are also preserved. Therefore, it can act as a criteria to assist PSACCF to carry out subsequent operations. Since resource preemption is forbidden in PSACCF, in other words, ongoing requests will not be forcibly dropped, so it's services whose actual CSARs lower than their target CSAR that should get preferred by SAC algorithm currently, we record them with a set $\mathcal{P}(t)$.

\begin{equation} \label{eq:target distance}
	TD_k(t) = \alpha^{\rm tar}_k(t) - \alpha_k(t)
\end{equation}

{\color{black}
PSACCF relies on the result of double sorts to determine the access order of the queues in $\mathcal{P}(t)$. {\bf Procedure~\ref{alg:p4 approach to the tar CSAR}} shows the operations. The elements in $\mathcal{P}(t)$ are arranged in descending order with respect to their resource efficiencies, and stably sorted again according to the distance between their actual CSARs and target CSARs, termed as $\boldsymbol{TD}(t)$ in Eq.~\eqref{eq:target distance}.

\begin{algorithm}
	\renewcommand{\algorithmcfname}{Procedure}
	\renewcommand{\thealgocf}{4}
	\caption{Approach to the target CSAR} \label{alg:p4 approach to the tar CSAR}
	Update $\boldsymbol{c}^{\rm idle}(t)$, $\boldsymbol{DR}(t)$, construct $\mathcal{P}(t)$\; \label{alg:P(t) begin}
	Descending sort $\mathcal{P}(t)$ with $\boldsymbol{RE}(t)$\;
	Stably descending sort $\mathcal{P}(t)$ with $\boldsymbol{TD}(t)$\;
	\For{$k \in \mathcal{P}(t)$ \label{alg:p4 approach begin}}{
		\If{$\boldsymbol{R} \cdot (\boldsymbol{a}(t) + \boldsymbol{e}_k) \leq \boldsymbol{c}^{\rm ava}(t)$}{
			$a_k(t)$++, $l_k(t)${-}{-}, update $\alpha_k(t)$\;
			\textbf{break}\; \label{alg:p4 approach end}
		}
	}
\end{algorithm}

Subsequently, requests belonging to queues in $\mathcal{P}(t)$ are tried to see whether the resource constraints still hold if one such request is accepted. If so, the top request in this queue is pulled to service and its CSAR is updated concurrently. Then the {\bf For Loop} interrupts as $\boldsymbol{\alpha}(t)$ has already changed, shown in Line~\ref{alg:p4 approach begin} to Line~\ref{alg:p4 approach end}. It is ensured in {\bf Procedure~\ref{alg:p4 approach to the tar CSAR}} that a new admitted request must be the one with the biggest gap to its target CSAR, or the one with a higher resource efficiency when several kinds of service possess the same $TD$. 
}

{\color{black}
The full pseudocode of our algorithm is given in Alg.~\ref{alg:PSACCF}. It takes the set of ongoing requests $\mathcal{O}(t)$, the queue length vector $\boldsymbol{l}(t)$ and the initial CSAR vector ${\bf \alpha}^{\rm I}(t)$ as inputs, outputs the admission decision $\boldsymbol{a}(t)$.

Line~\ref{alg:PSACCF init} mainly initializes variables and counts the loss of requests caused by impatient behaviors, thereby updating the queue length, available resources, and the dominant resource labels in Line~\ref{alg:PSACCF update after init}.
}

%Once the admission decision about $\mathcal{P}(t)$ successfully yields a new admitted request, the testing procedure should be interrupted as $\boldsymbol{\alpha}(t)$ has already updated. Therefore, the fairness indicator should be reevaluated and target CSARs should also be recalculated to guide subsequent admission actions. The fairness enhancement process repeats until no more optimizations can be done to $\boldsymbol{\alpha}(t)$.

\begin{algorithm}[!th]
	\renewcommand{\thealgocf}{1}
	\caption{PSACCF} \label{alg:PSACCF}
	\SetKwInOut{Input}{Input}
	\SetKwInOut{Output}{Output}
	
	\Input{1) Set of ongoing requests $\mathcal{O}(t)$; 2) Length of queues $\boldsymbol{l}(t)$; 3) Initial CSAR vector $\boldsymbol{\alpha}^{\rm I}(t)$.}
	\Output{Admission decision $\boldsymbol{a}(t)$.}
	Initiate $\boldsymbol{a}(t)=0$, $\boldsymbol{\alpha}(t)=\boldsymbol{\alpha}^{\rm I}(t)$, count $\boldsymbol{d}(t)$ and $\boldsymbol{b}(t)$\; \label{alg:PSACCF init}
	Update $\boldsymbol{l}(t)$, $\boldsymbol{c}^{\rm ava}(t)$ and $\boldsymbol{DR}(t)$\; \label{alg:PSACCF update after init}
	\tcp{priority amendment}
	Execute \textbf{Procedure~\ref{alg:p1 fix priority}}\; \label{alg:PSACCF p1}
	\If{$\mathcal{I}(\boldsymbol{\alpha}(t))==1$ \label{alg:PSACCF check priority}}{
		\tcp{fairness enhancement}
		\While{true}{
			\tcp{check the degree of fairness}
			Execute \textbf{Procedure~\ref{alg:check fairnes}}\; \label{alg:PSACCF p2}
			\tcp{calculate the target CSAR}
			Execute \textbf{Procedure~\ref{alg:p3 calculate the tar CSAR}}\; \label{alg:PSACCF p3}
			\tcp{approach to the target CSAR}
			Execute \textbf{Procedure~\ref{alg:p4 approach to the tar CSAR}}\; \label{alg:PSACCF p4}
			\If{$\boldsymbol{\alpha}(t)$ remains unchanged after {\rm \textbf{Procedure~\ref{alg:p4 approach to the tar CSAR}}}}{
				\textbf{break}\;
			}
		}
	}
	\textbf{return} $\boldsymbol{a}(t)$\;
\end{algorithm}

{\color{black}
Line~\ref{alg:PSACCF p1} checks the state of priority and tries to correct violations when there has. The priority requirement is twice inspected in Line~\ref{alg:PSACCF check priority}, if the amendment in Line~\ref{alg:PSACCF p1} failed, it must be due to the shortage of resources, meaning that it is impossible to improve fairness, so the algorithm ends. Otherwise, PSACCF enters the phase of fairness enhancement.

The fairness degree is checked in Line~\ref{alg:PSACCF p2} and a second strengthen for priority requirement is carried out here if the fairness is high enough. Then the calculation of target CSARs is executed in Line~\ref{alg:PSACCF p3}, closely followed by {\bf Procedure~\ref{alg:p4 approach to the tar CSAR}}, which is responsible for pushing the actual CSARs toward the target CSARs. Each time a new request is accepted, $\boldsymbol{\alpha}(t)$ will change accordingly, so does $f_{J}(\boldsymbol{\alpha}(t))$. Therefore, {\bf Procedure~\ref{alg:check fairnes}} to {\bf Procedure~\ref{alg:p4 approach to the tar CSAR}} need to be executed repeatedly until no more optimizations can be made on $\boldsymbol{\alpha}(t)$.
}

\subsection{Resource scheduling}

An operation that is tightly coupled to admission control is resource allocation. In our scenario, where elastic and inelastic services coexist, requests that originate from different categories are treated differently.

Inelastic requests are granted exactly the amount they demand once admitted and remain unchanged during the whole lifetime. It takes more effort when it comes to elastic requests. Newly admitted elastic requests also receive the requested quantity upon admission but still can be adjusted later, if there are residual resources left when PSACCF is done, they will be evenly shared by ongoing elastic requests with the highest priority level. On the one hand, this allows for better utilization of resources, which echoes the third objective. On the other hand, it enables the effect of relocating the current idle resources to future time slots. More specifically, in the next time slot, over-provisioned requests can be reallocated with the minimum resources that ensure on-time completions, saving more space to accommodate future requests, which looks as if the previous resources were delayed to the subsequent time slot.

\subsection{Complexity analysis}
{\color{black}
This subsection explains the complexity of PSACCF. The time costs of {\bf Procedure~\ref{alg:p1 fix priority}} to {\bf Procedure~\ref{alg:p4 approach to the tar CSAR}} are elaborated at first, based on which the time consumption of PSACCF is deduced.

In {\bf Procedure~\ref{alg:p1 fix priority}}, Line~\ref{alg:p1 construct violation set} takes $O(K^2)$, the {\bf for Loop}  operates at most $K$ times, each takes $O(N)$. The outer {\bf while Loop} stops when the priority requirement is met or no resources are available, since the total capacity and demand vectors are fixed, its upper bound is limited to the maximum supportable number of requests when resources are exclusively consumed by one kind of service. We use a constant $C$, independent of $K$ and $N$, to term the bound. Then, {\bf Procedure~\ref{alg:p1 fix priority}} takes $O(C \cdot K^2 + C \cdot K \cdot N)$. Similar analysis can be applied to {\bf Procedure~\ref{alg:check fairnes}}, it decreases to $O(C \cdot K \cdot N)$ for the absence of sort operation. As for {\bf Procedure~\ref{alg:p3 calculate the tar CSAR}} and {\bf Procedure~\ref{alg:p4 approach to the tar CSAR}}, it's easy to understand that their time complexity are $O(K)$ and $O(K^2 + K \cdot N)$ respectively. 

In Alg.~\ref{alg:PSACCF}, the Line~\ref{alg:PSACCF init} takes $O(K)$ and Line~\ref{alg:PSACCF update after init} takes $O(K \cdot N)$. Priority amendment takes $O(C \cdot K^2 + C \cdot K \cdot N)$. The {\bf While Loop} will interrupt when fairness is no longer optimizable or resources are saturated, which means it also can not exceed $C$ cycles, which means the time complexity of fairness enhancement is $O(C^2 \cdot K \cdot N + C \cdot K + C \cdot K^2 +C \cdot K \cdot N)$. Thus, the overall time complexity of PSACCF is $O\left((C+1) \cdot K + 2C \cdot K^2 + (C^2 + 2C +1) \cdot K \cdot N\right)$. Considering that $C$ is nothing to do with $K$ and $N$, so the final result is $O(K^2 + K \cdot N)$.
}

%% file: section/sec06.PerformanceEvaluation.tex
\section{Performance Evaluation} \label{performance evaluation}

%In this section, numerical results are presented in the form of figures to demonstrate the advantages of PSACCF.
% Explanations are attached to each graph to help the understanding of some interesting phenomena and reveal some hidden laws.

\subsection{Simulation settings}

\textcolor{black}{Simulations are performed on a laptop with an i5-7300HQ CPU and 16G RAM, the software environment is Windows~10 Professional and C++ 8.1.0.}

Two kinds of services are considered during the simulation and both can generate handoff requests, i.e., $K=4$. Each request is assumed to require two types of resources, i.e., $N=2$. Similar as the normalized resource vector in~\cite{han2020multiservice}, the first kind of service is elastic with demand vector $(0.035,\, 0.03)^T$, the second is inelastic with demand vector $(0.016,\, 0.02)^T$, and the resource capacity is $(2.0,\, 2.0)^T$. We use Service-1 to Service-4 to represent these four services, the first two denote new requests and the last two denote handoff requests, which have higher priorities.

\textcolor{black}{We assume that the different requests have their own arrival rates, which are factored against a base arrival rate $\lambda_0$ by [1.2, 1.5, 0.6, 0.75] respectively. So do the lifetime and hold time parameters, their factors are [0.6, 1.0, 0.3, 0.5] and [1.0, 0.8, 0.3, 0.24]. This setting implies that handoff requests usually have a shorter duration and need to be dealt with more timely.}

\begin{table}[!t]
	\caption{Parameter Settings} \label{tab:para}
	\centering
	\renewcommand{\arraystretch}{1.5}
	\begin{tabular}{cccc}
		\hline
		symbol & value & symbol & value \\
		\hline
		$K$ & 4 & $N$ & 2 \\
%		\hline
		$T^{\rm L}_0$ & 20.0 & $T^{\rm H}_0$ & 10.0 \\
%		\hline
		$T^{\rm L}_1$ & $0.6 \cdot T^{\rm L}_0$ & $T^{\rm H}_1$ & $1.0 \cdot T^{\rm H}_0$ \\
%		\hline
		$T^{\rm L}_2$ & $1.0 \cdot T^{\rm L}_0$ & $T^{\rm H}_2$ & $0.8 \cdot T^{\rm H}_0$ \\
%		\hline
		$T^{\rm L}_3$ & $0.3 \cdot T^{\rm L}_0$ & $T^{\rm H}_3$ & $0.3 \cdot T^{\rm H}_0$ \\
%		\hline
		$T^{\rm L}_4$ & $0.5 \cdot T^{\rm L}_0$ & $T^{\rm H}_4$ & $0.24 \cdot T^{\rm H}_0$ \\
%		\hline
		$p_0$ & 1.0 & $\beta_0$ & 0.02 \\
%		\hline
		$p_k$ & $(1+{\rm log}(k)) \cdot p_0$ & $\beta_k$ & $\beta_0/(1+k)$ \\
%		\hline
		$pr$ & 1.5 & $\boldsymbol{w}$ & $(1,1,\dots,1)^T$ \\
%		\hline
		$\lambda_1$ & $1.2 \cdot \lambda_0$ & $\boldsymbol{r}_1$ & $(0.035,\, 0.03)^T$ \\
%		\hline
		$\lambda_2$ & $1.5 \cdot \lambda_0$ & $\boldsymbol{r}_2$ & $(0.016,\, 0.02)^T$ \\
%		\hline
		$\lambda_3$ & $0.6 \cdot \lambda_0$ & $\boldsymbol{r}_3$ & $(0.035,\, 0.03)^T$ \\
%		\hline
		$\lambda_4$ & $0.75 \cdot \lambda_0$ & $\boldsymbol{r}_4$ & $(0.016,\, 0.02)^T$ \\
%		\hline
		$\lambda_0$ & $\{ 5,\, 7,\, 9,\, 11,\, 13,\, 15 \}$ & $\boldsymbol{c}$ & $(2.0,\, 2.0)^T$ \\
%		\hline
		$\epsilon$ & $\{ 0.05,\, 0.10,\, 0.15 \}$ & $\varphi$ & $\{ 0.85,\, 0.90,\, 0.95,\, 1.0 \}$ \\
		\hline
	\end{tabular}
\end{table}

%\begin{table}[!h]
%	\caption{Parameter Settings} \label{tab:para}
%	\centering
%	\renewcommand{\arraystretch}{1.5}
%	\begin{tabular}{cccccc}
%		\hline
%		symbol & value & symbol & value & symbol & value \\
%		\hline
%		$K$ & 4 & $N$ & 2 & $\lambda_0$ & $\{ 5,\, 7,\, 9,\, 11,\, 13,\, 15 \}$\\
%		%		\hline
%		$T^{\rm L}_0$ & 20.0 & $T^{\rm H}_0$ & 10.0 & $\lambda_1$ & $1.2 \cdot \lambda_0$ \\
%		%		\hline
%		$T^{\rm L}_1$ & $0.6 \cdot T^{\rm L}_0$ & $T^{\rm H}_1$ & $1.0 \cdot T^{\rm H}_0$ & $\lambda_2$ & $1.5 \cdot \lambda_0$ \\
%		%		\hline
%		$T^{\rm L}_2$ & $1.0 \cdot T^{\rm L}_0$ & $T^{\rm H}_2$ & $0.8 \cdot T^{\rm H}_0$ & $\lambda_3$ & $0.6 \cdot \lambda_0$ \\
%		%		\hline
%		$T^{\rm L}_3$ & $0.3 \cdot T^{\rm L}_0$ & $T^{\rm H}_3$ & $0.3 \cdot T^{\rm H}_0$ & $\lambda_4$ & $0.75 \cdot \lambda_0$ \\
%		%		\hline
%		$T^{\rm L}_4$ & $0.5 \cdot T^{\rm L}_0$ & $T^{\rm H}_4$ & $0.24 \cdot T^{\rm H}_0$ & $\boldsymbol{c}$ & $(2.0,\, 2.0)^T$ \\
%		%		\hline
%		$p_0$ & 1.0 & $\beta_0$ & 0.02 & $\boldsymbol{r}_1$ & $(0.035,\, 0.03)^T$ \\
%		%		\hline
%		$p_k$ & $(1+{\rm log}(k)) \cdot p_0$ & $\beta_k$ & $\beta_0/(1+k)$ & $\boldsymbol{r}_2$ & $(0.016,\, 0.02)^T$ \\
%		%		\hline
%		$pr$ & 1.5 & $\boldsymbol{w}$ & $(1,1,\dots,1)^T$ & $\boldsymbol{r}_3$ & $(0.035,\, 0.03)^T$  \\
%		%		\hline
%		$\epsilon$ & $\{ 0.05,\, 0.10,\, 0.15 \}$ & $\varphi$ & $\{ 0.85,\, 0.90,\, 0.95,\, 1.0 \}$ & $\boldsymbol{r}_4$ & $(0.016,\, 0.02)^T$ \\
%		\hline
%	\end{tabular}
%\end{table}

\textcolor{black}{The parameter $\beta_k$ decreases as the priority level $k$ rises, indicating a higher willingness for a request to join a queue.} So do the service price $p_k$, i.e., the fees a request needs to pay for each time slot. Certain values are summarized in Tab.~\ref{tab:para}. Notice that failure to complete ongoing requests yields a penalty, we represent this by a penalty ratio, termed as $pr$, which is set to 1.5.

For convenience of simulation, $\boldsymbol{w}$ is fixed to $\boldsymbol{1}$. While, different $\lambda_0$, $\epsilon$ and $\varphi$ are tested to mining potential abilities of PSACCF, values used are also recorded in Tab.~\ref{tab:para}.

\subsection{Self-evaluation}

{\color{black}
First of all, we fix $\epsilon$ and $\varphi$ to test the performance of PSACCF under different request arrival rates, results are shown in Fig.~\ref{fig:PSACCF CSAR lambda}. The curves correspond to the CSARs of each service queue during 1000 time slots. As the load increases, CSARs of all four services decrease, because balk and renege become more serious. However, curves smooth out quickly regardless of $\lambda_0$, and the gap between them remains basically unchanged, which reflects the effect of $\epsilon$ and $\varphi$.

\begin{figure}[!h]
	\centering
	\subfloat[$\lambda_0=5$]{
		\centering
		\includegraphics[scale=0.4]{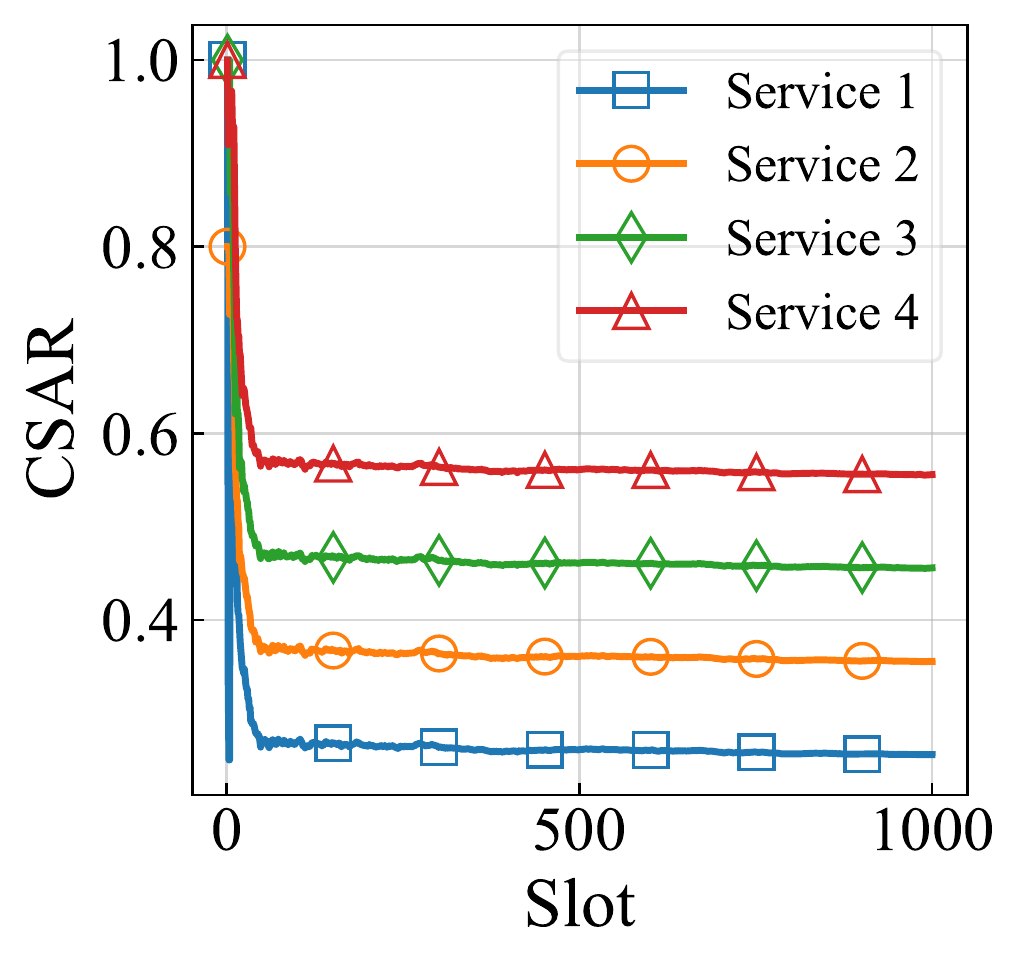}
	}
	%	\subfloat[$\lambda_0=7$]{
		%		\centering
		%		\includegraphics[scale=0.4]{PSACCF_Admit_Ratio_beta=0.02_lambda=7.00_epsilon=0.10_fair=1.00}
		%	}
	\subfloat[$\lambda_0=9$]{
		\centering
		\includegraphics[scale=0.4]{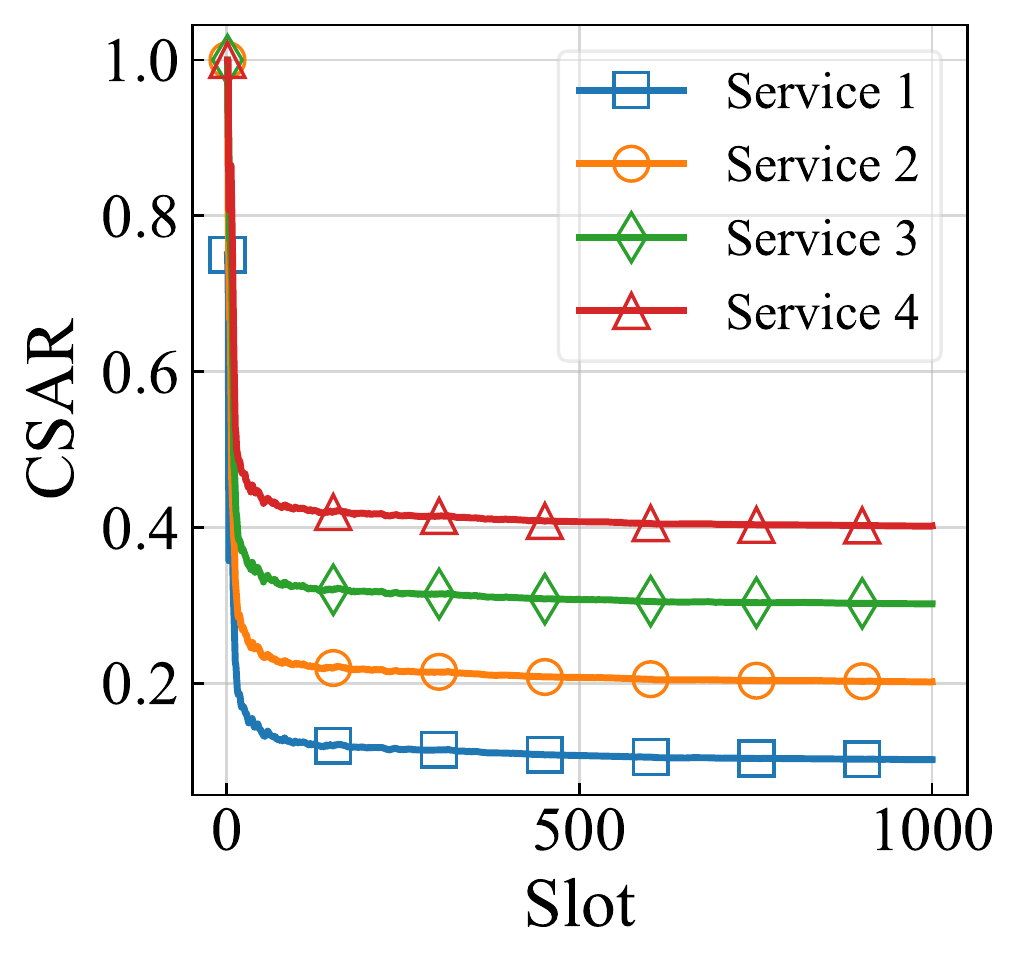} \label{fig:PSACCF CSAR lambda=9}
	}
	%	\subfloat[$\lambda_0=11$]{
		%		\centering
		%		\includegraphics[scale=0.4]{PSACCF_Admit_Ratio_beta=0.02_lambda=11.00_epsilon=0.10_fair=1.00}
		%	}
	%	\subfloat[$\lambda_0=13$]{
		%		\centering
		%		\includegraphics[scale=0.4]{PSACCF_Admit_Ratio_beta=0.02_lambda=13.00_epsilon=0.10_fair=1.00}
		%	}
	%	\subfloat[$\lambda_0=15$]{
		%		\centering
		%		\includegraphics[scale=0.4]{PSACCF_Admit_Ratio_beta=0.02_lambda=15.00_epsilon=0.10_fair=1.00}
		%	}
	
	\caption{CSAR of PSACCF when $\epsilon=0.1, \varphi=1$} \label{fig:PSACCF CSAR lambda}
\end{figure}

\begin{figure}[!h]
	\centering
	\subfloat[$\epsilon=0.1, \varphi=0.85$]{
		\includegraphics[scale=0.4]{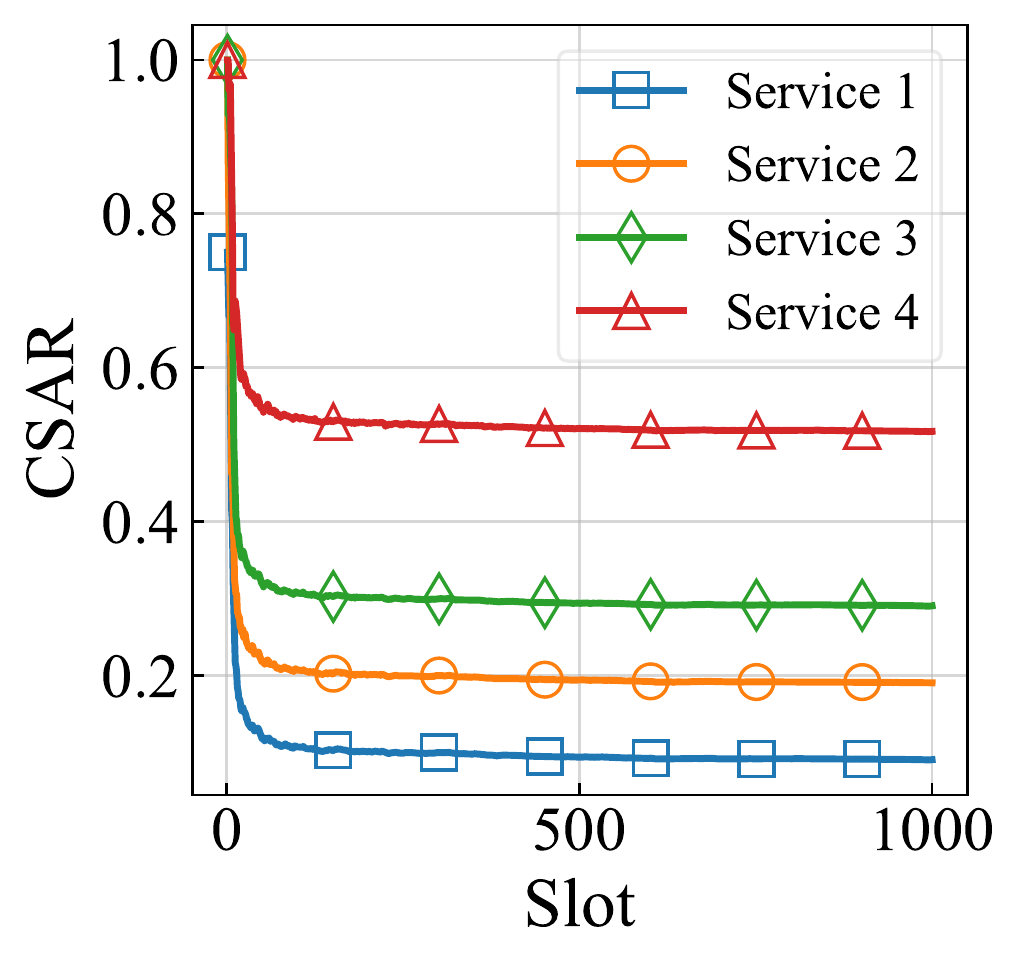}
		\label{fig:PSACCF CSAR l=9 e=0.1 v=0.85}
	}
	\subfloat[$\epsilon=0.15, \varphi=1$]{
		\includegraphics[scale=0.4]{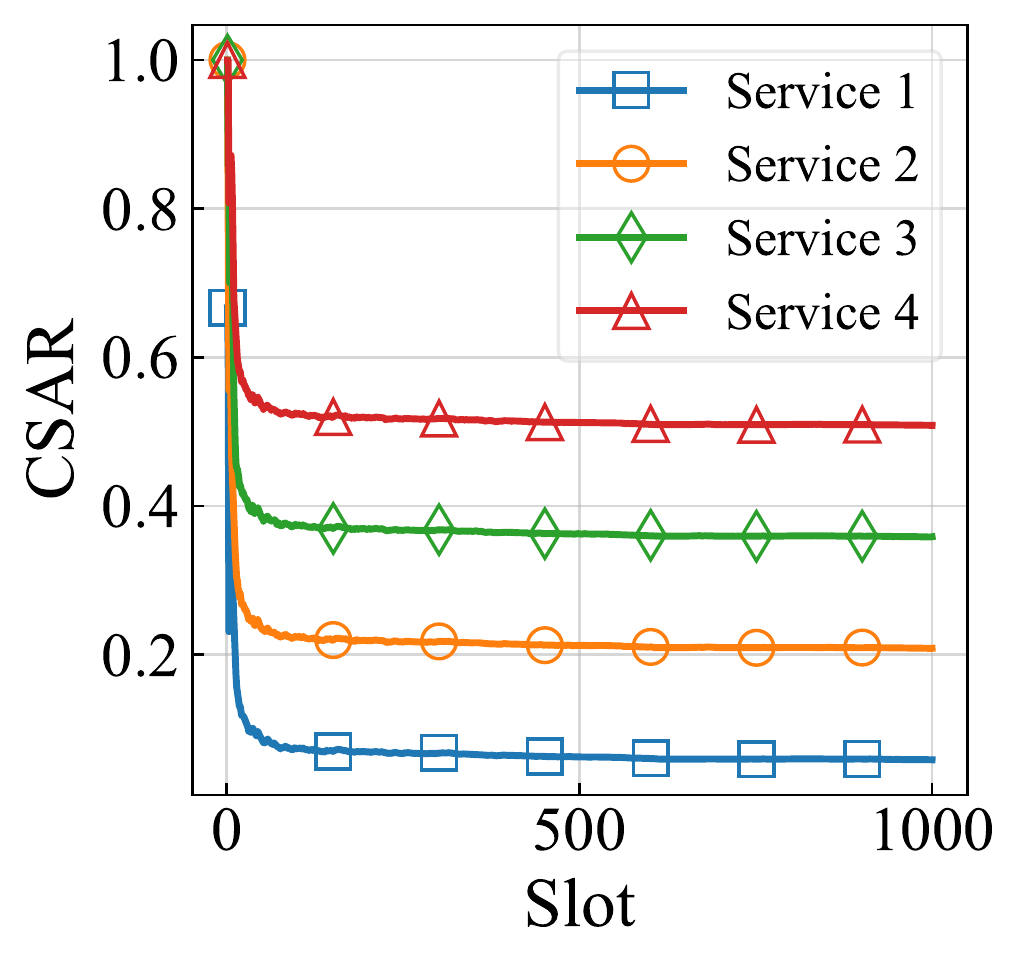}
		\label{fig:PSACCF CSAR l=9 e=0.15 v=1}
	}
	\caption{CSAR of PSACCF when $\lambda_0=9$} \label{fig:PSACCF CSAR l=9}
\end{figure}

Fig.~\ref{fig:PSACCF CSAR l=9} further explains how $\epsilon$ and $\varphi$ influence the CSARs by inspecting other combinations of them when $\lambda_0$ is fixed to 9. In more detail, Fig.~\ref{fig:PSACCF CSAR l=9 e=0.1 v=0.85} relaxes $\varphi$ to 0.85 as compared with Fig.~\ref{fig:PSACCF CSAR lambda=9}. Each moment the fairness threshold is reached, services with higher priority will receive more attention from PSACCF, making an expanded gap between Service-4 and Service-3 than others. While, Fig.~\ref{fig:PSACCF CSAR l=9 e=0.15 v=1} adopts a bigger $\epsilon$, yielding larger absolute gaps of CSAR, but the uniformity of them is similar to that in Fig.~\ref{fig:PSACCF CSAR lambda=9} for a consistent $\varphi$.

It can be concluded from Fig.~\ref{fig:PSACCF CSAR lambda} and Fig.~\ref{fig:PSACCF CSAR l=9} that the CSAR of high priority can be promoted at the expense of depressing the CSAR of low priority, by raising $\epsilon$ or suppressing $\varphi$, and vice versa. It looks like there are more resources reserved for requests with higher priority. What's interesting is that the amount of resources set aside is no longer immutable, but will adjust automatically depending on the request arrival rates, which makes PSACCF expertise in scheduling resources flexibly and then contributes to better resource utilization.  

%To show more clearly what $\epsilon$ and $\varphi$ do, additional data are presented in Fig.~\ref{fig:PSACCF CSAR l=9}, where $\lambda_0$ is fixed to 9, but different combinations of $\epsilon$ and $\varphi$ are tested. Comparing Fig.~\ref{fig:PSACCF CSAR l=9 e=0.1 v=0.85} with Fig.~\ref{fig:PSACCF CSAR lambda=9}, since the fairness threshold is relaxed, higher priority services will receive more attention when $\varphi$ is reached, so the gap between fourth and third services is much wider than others. While, in Fig.~\ref{fig:PSACCF CSAR l=9 e=0.15 v=1}, the uniformity of curve gaps is similar to that in Fig.~\ref{fig:PSACCF CSAR lambda=9}, but the size is enlarged due to a higher $\epsilon$. Putting three figures together, CSAR of high priority service can be promoted at the expense of depressing the CSAR of low priority service by raising $\epsilon$ or suppressing $\varphi$, and vice versa. From the results, it looks as if there are more resources set aside for high priority requests. And what interesting is the amount of resources reserved is no longer immutable, but will automatically adjust itself depending on the arrival rates of requests, which contributes to a better resource utilization.

\begin{figure}[!th]
	\centering
	\subfloat[$\lambda_0=7$]{
		\centering
		\includegraphics[scale=0.4]{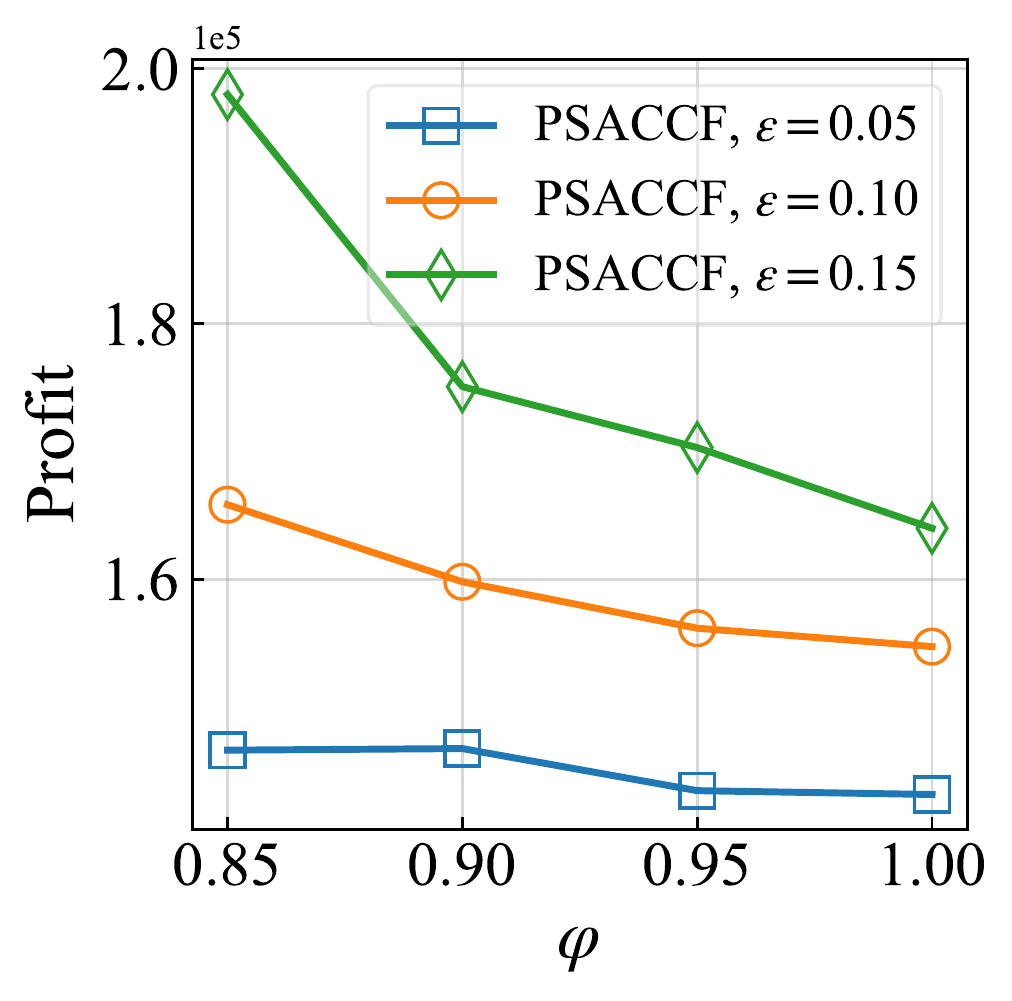}
	}
	\subfloat[$\lambda_0=13$]{
		\centering
		\includegraphics[scale=0.4]{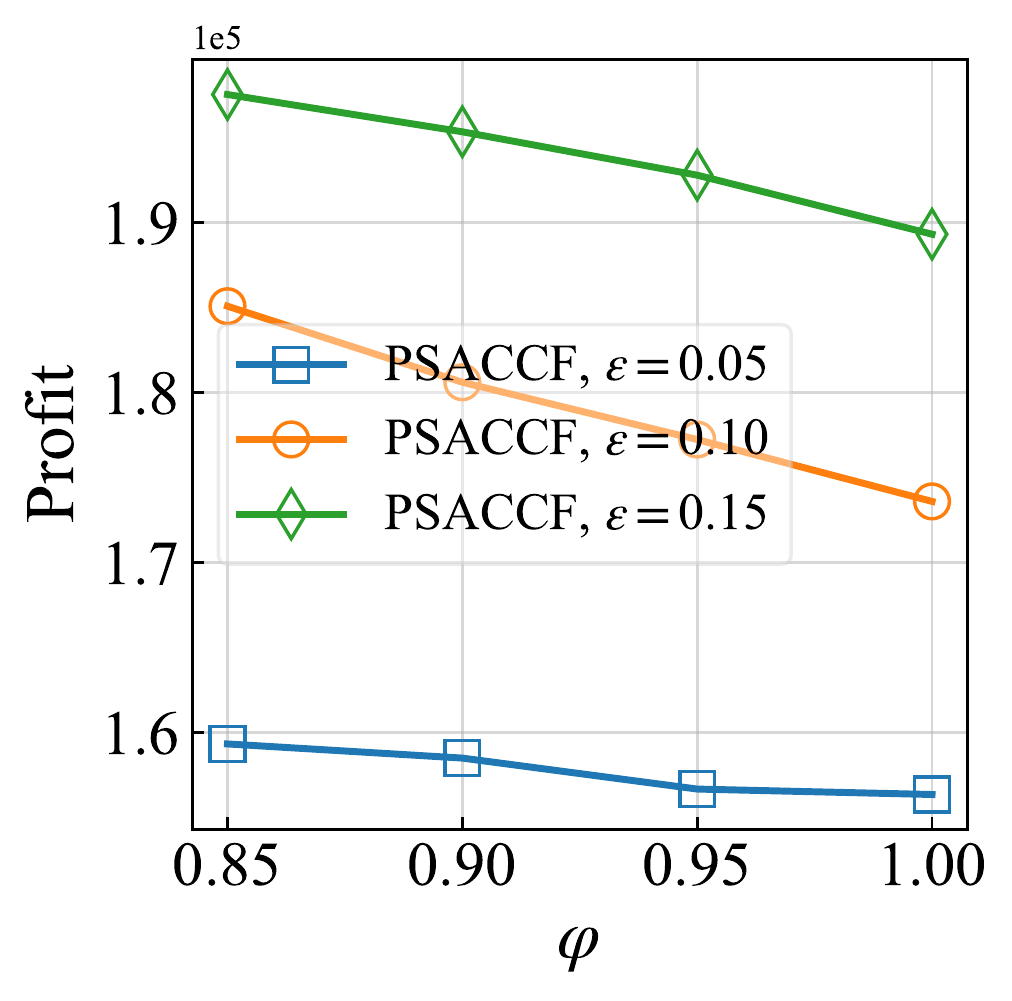}
	}
	
	\caption{Profit of PSACCF under different workload} \label{fig:PSACCF profit varphi}
\end{figure}

All combinations of $\epsilon$ and $\varphi$ are tested in Fig.~\ref{fig:PSACCF profit varphi} to observe the response of profit earned by the SP. Phenomena exhibited comply with those in Fig.~\ref{fig:PSACCF CSAR lambda} and Fig.~\ref{fig:PSACCF CSAR l=9}. When the uniformity of CSAR gaps gets more rigorous, requests with higher priority are less admitted, leading to a lower total profit. On the contrary, larger $\epsilon$ makes the CSAR curves sparser, which is in favor of high priority requests, thus generating more profitable admission decisions because the principle that higher priority affords higher service fees is assumed in our work.

%We enumerate all the combinations of $\epsilon$ and $\varphi$ in Fig.~\ref{fig:PSACCF profit varphi}, where the profit earned by SP after 1000 time slots is observed. Phenomena here comply with those appeared above. When $\varphi$ increases, meaning the uniformity of CSAR gaps gets more rigorous, less high priority requests is admitted, leading to a lower total profits. To the contrary, larger $\epsilon$ makes the four CSAR curves looser, which is in favor of high priority requests, and generates more profitable admission decisions when the principle that higher priority affords higher service fees is assumed.

Figures listed in this subsection demonstrate that PSACCF not only persists trustworthy stability when the level of workload varies, but also offers strong flexibility and adaptability for the SP to cope with diverse situations just by tuning $\epsilon$ and $\varphi$.
}

\subsection{Comparison of Algorithms}

%In this subsection, we compare the performance of PSACCF with two existing admission policies proposed in \cite{perveen2021dynamic} and \cite{kushchazli2021model} on each objective under the same environment. To begin with, brief descriptions about these latter two algorithms are presented.

{\color{black}
Except for PSACCF, algorithms proposed in \cite{perveen2021dynamic} and \cite{kushchazli2021model} are evaluated  as comparisons below. The former, denoted as MHPF, groups the received requests into clusters according to their characteristics, and then accesses them in the order of High Priority First using the priority of the cluster, the newly accepted requests cannot interfere with the resources occupied by other in-service requests. Higher Priority First is still persisted in the latter, denoted as AHPF. But a more aggressive resource scheduling scheme is adopted, that is new requests are permitted to grab resources from ongoing services as long as they are more privileged. Since only two priority levels are considered in \cite{kushchazli2021model}, we slightly expand it to our scenario, where multiple levels coexist, by setting the preemption order starting with the lowest priority.

%The admission algorithm in \cite{perveen2021dynamic} first clusters the received requests by characteristics, and then accepts them according to the priority of the cluster in the principle of higher priority first, the newly accessed requests cannot interfere with the resources of other already in-service requests, we denote it with MHPF in short. While, in \cite{kushchazli2021model}, higher priority first is still persisted, but a more aggressive resource scheduling scheme is adopted, new requests are permitted to grab resources from ongoing requests as long as the former is more privileged. Since only two priority levels are considered in \cite{kushchazli2021model}, we slightly expand it to our scenario where multiple levels exist by setting the preemption order starting with low priority requests, we call it AHPF shortly.

\begin{figure}[!h]
	\centering
	\subfloat[$\lambda_0=5$]{
		\centering
		\includegraphics[scale=0.4]{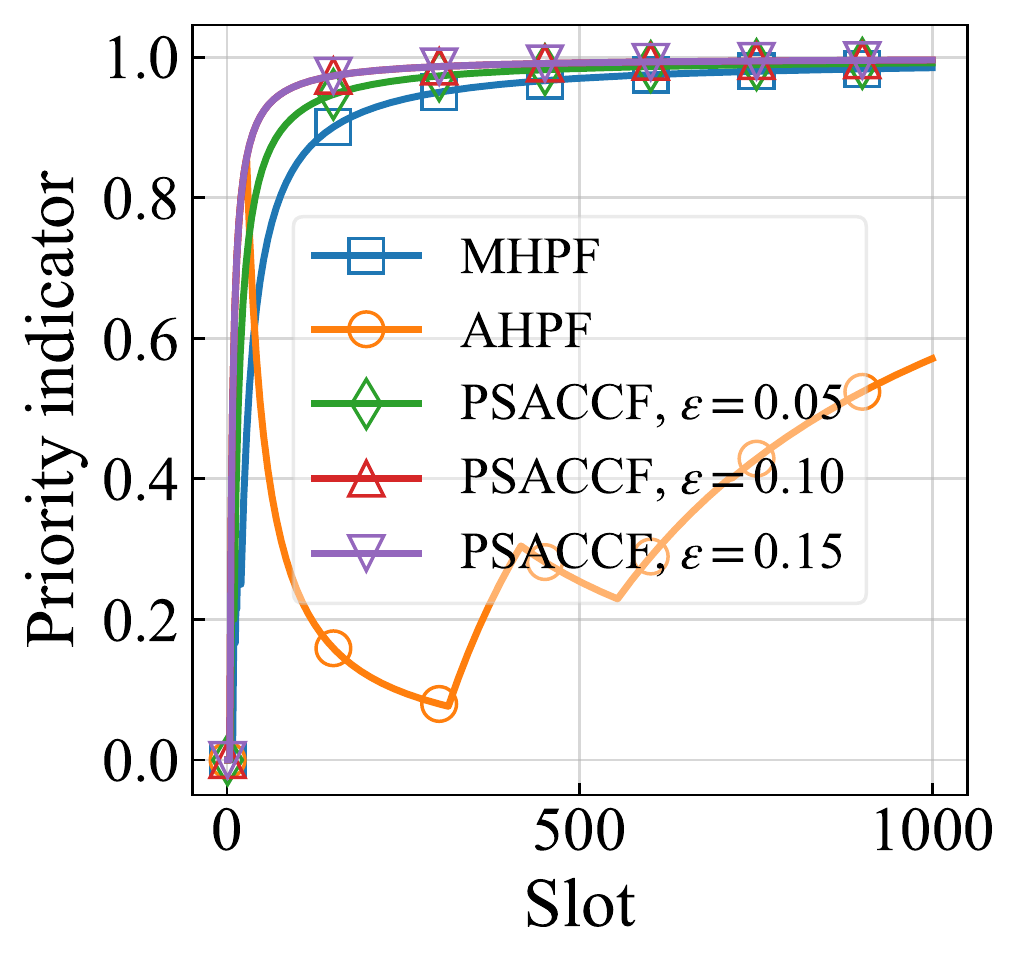}
		\label{fig:Priority indicator lambda=5}
	}
	\subfloat[$\lambda_0=15$]{
		\centering
		\includegraphics[scale=0.4]{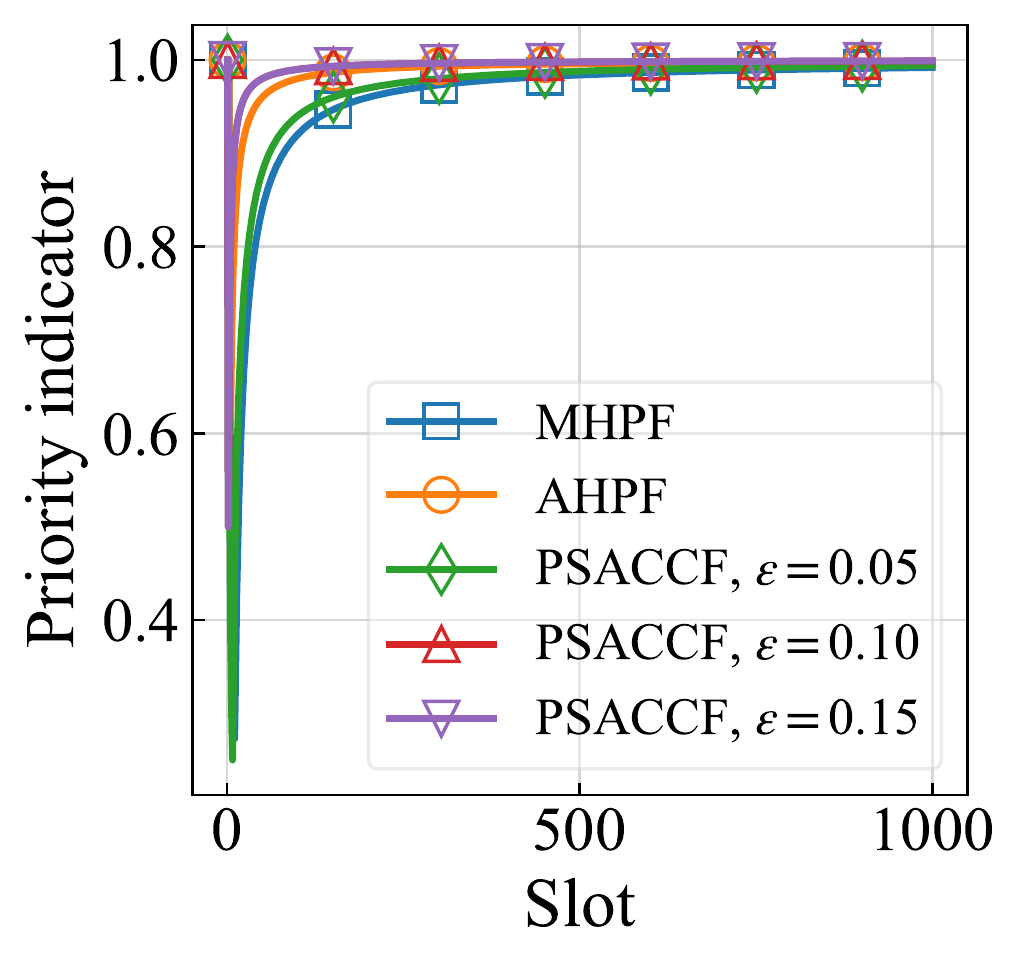}
	}
	
	\caption{Priority indicator of three algorithms when $\varphi=1$} \label{fig:Priority indicator}
\end{figure}

\subsubsection{Priority indicator} \ \par

The priority objective of the three algorithms is shown in Fig.~\ref{fig:Priority indicator} by observing the priority indicator defined in Eq.~\eqref{eq:CSAR} for 1000 slots. The fairness threshold $\varphi$ is set to 1.0, and in each subgraph, PSACCF splits into three curves, each with a specific $\epsilon$. In the beginning, the priority indicator is low, because Eq.~\eqref{eq:CSAR} is vulnerable to rational behaviors when the cumulative number of accepted requests is small. Then, the curves of PSACCF and MHPF rise sharply in the early stage, during which PSACCF better performs than MHPF, especially under a higher $\epsilon$. Eventually this indicator approaches each other for both PSACCF and MHPF.

%To evaluate the priority objective of three algorithms, we fix the fairness threshold of PSACCF to 1.0 and observe the priority indicator for 1000 slots under different $\lambda_0$, the results are shown in Fig.~\ref{fig:Priority indicator}. In each subgraph, PSACCF is divided into three curves, each with a specific $\epsilon$. At the beginning, the priority indicator is small, because Eq.~\eqref{eq:CSAR} is sensitive to rational behaviors when the cumulative number of requests accepted is low. Then, curves of PSACCF and MHPF rise sharply in the early stage, during which PSACCF better performs than MHPF especially with a higher $\epsilon$. And eventually this indicator is close to each other for both PSACCF and MHPF.

When the workload gets heavier, AHPF shows a similar pattern as others and outperforms PSACCF (with small $\epsilon$) and MHPF, which is in line with intuition and expectation. Nevertheless, when $\lambda_0$ takes a small value, the curve of AHPF becomes wired, it neither betters any algorithm nor remains stable. The reason is that resource seizure only occurs in the first two kinds of service to accommodate requests for the last two when the workload is light. But the arrival rate of Service-4 is higher than Service-3, which means more balks will occur correspondingly and results in a higher CSAR for the third than the fourth sometimes (see Fig.~\ref{fig:CSAR AHPF lambda=5}). Then, a violation of the priority requirement happens. When $\lambda_0$ becomes larger, preemption continues to spread to a higher level (see Fig.~\ref{fig:CSAR AHPF lambda=15}), the priority requirement gets back on track again, so the indicator improves. Evidence can be found in Fig.~\ref{fig:CSAR AHPF}, where the CSAR curves generated by AHPF of all services are shown.

\begin{figure}[!h]
	\centering
	\subfloat[$\lambda_0=5$]{
		\includegraphics[scale=0.4]{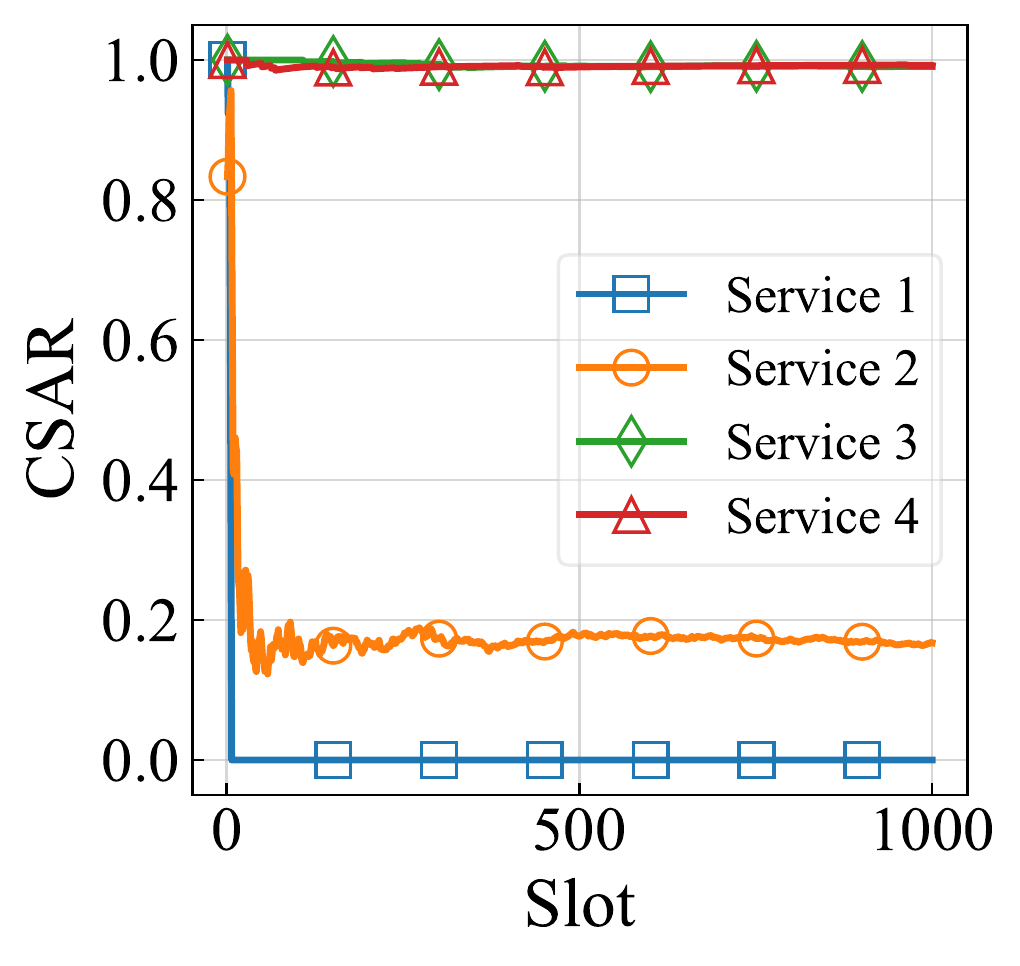}
		\label{fig:CSAR AHPF lambda=5}
	}
	\subfloat[$\lambda_0=15$]{
		\includegraphics[scale=0.4]{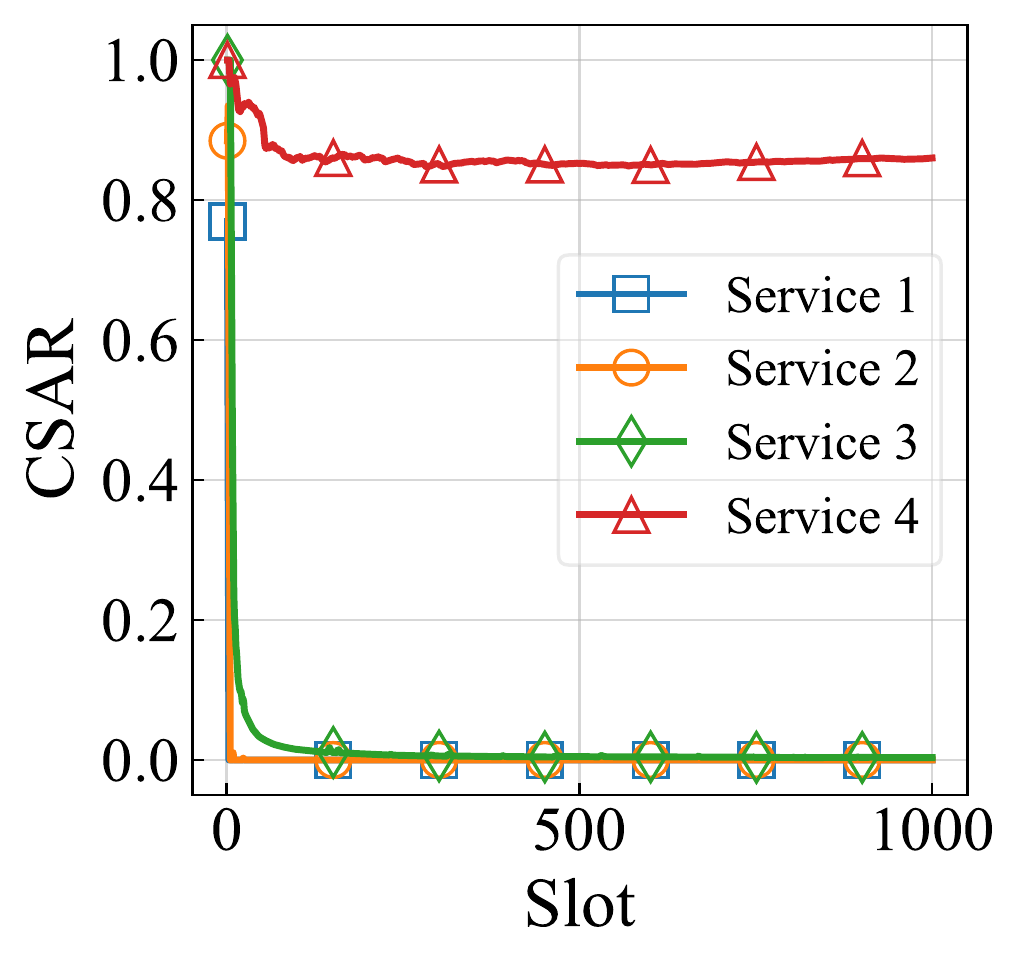}
		\label{fig:CSAR AHPF lambda=15}
	}
	
	\caption{CSAR of AHPF} \label{fig:CSAR AHPF}
\end{figure}

It can be concluded from Fig.~\ref{fig:Priority indicator} and Fig.~\ref{fig:CSAR AHPF} that PSACCF can achieve almost the same priority performance as the comparisons, and even a faster climb speed in the growth phase (under a bigger $\epsilon$). What's more, PSACCF performs smoothly against changes in the workload rather than shaking badly as AHPF, which reveals its advantages in the presence of balk behavior.

%It can be concluded from Fig.~\ref{fig:Priority indicator} and Fig.~\ref{fig:CSAR AHPF} that both PSACCF and MHPF are able to maintain CSAR differences more consistently over time, and by adjusting $\epsilon$, the priority indicator curve of PSACCF can climb faster. While, the performance of AHPF highly depends on the degree of load, i.e., it behaves extremely unstably in the presence of rational behaviors.

\begin{figure}[!th]
	\centering
	\subfloat[$\lambda_0=5$]{
		\centering
		\includegraphics[scale=0.4]{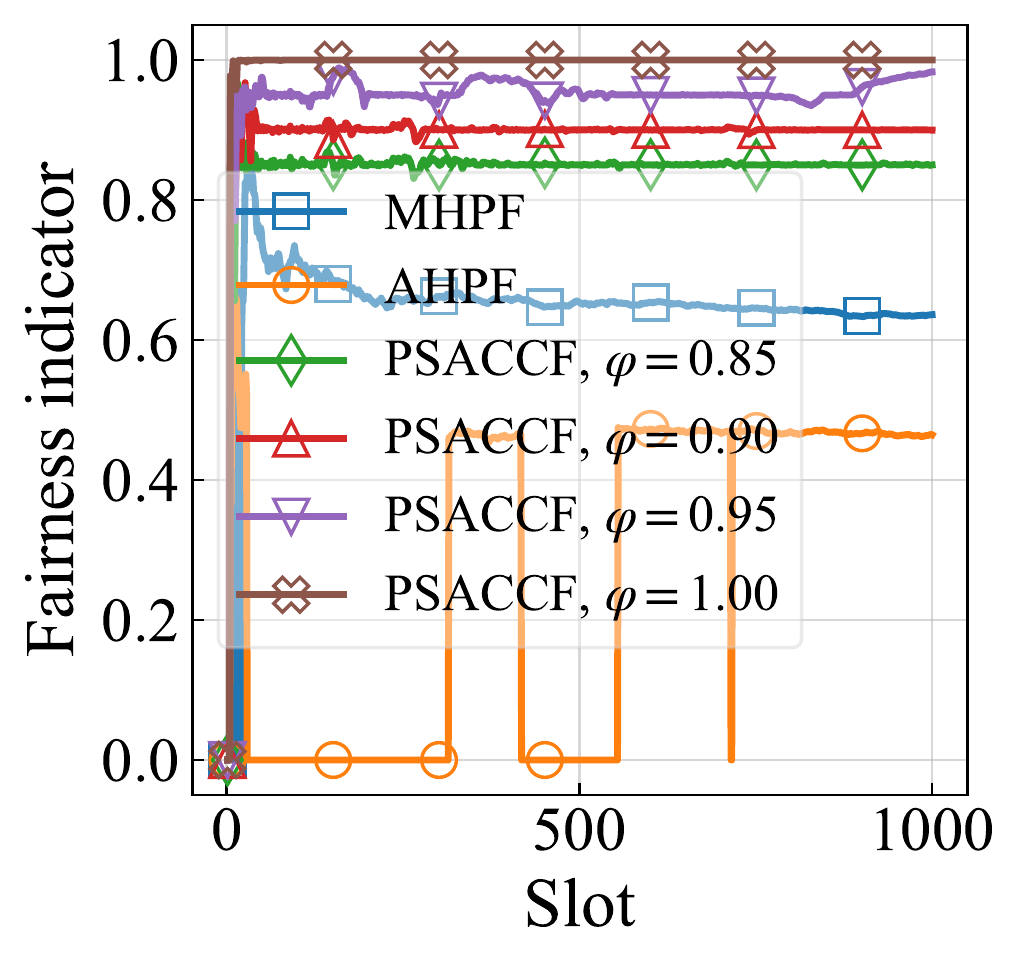}
	}
	\subfloat[$\lambda_0=15$]{
		\centering
		\includegraphics[scale=0.4]{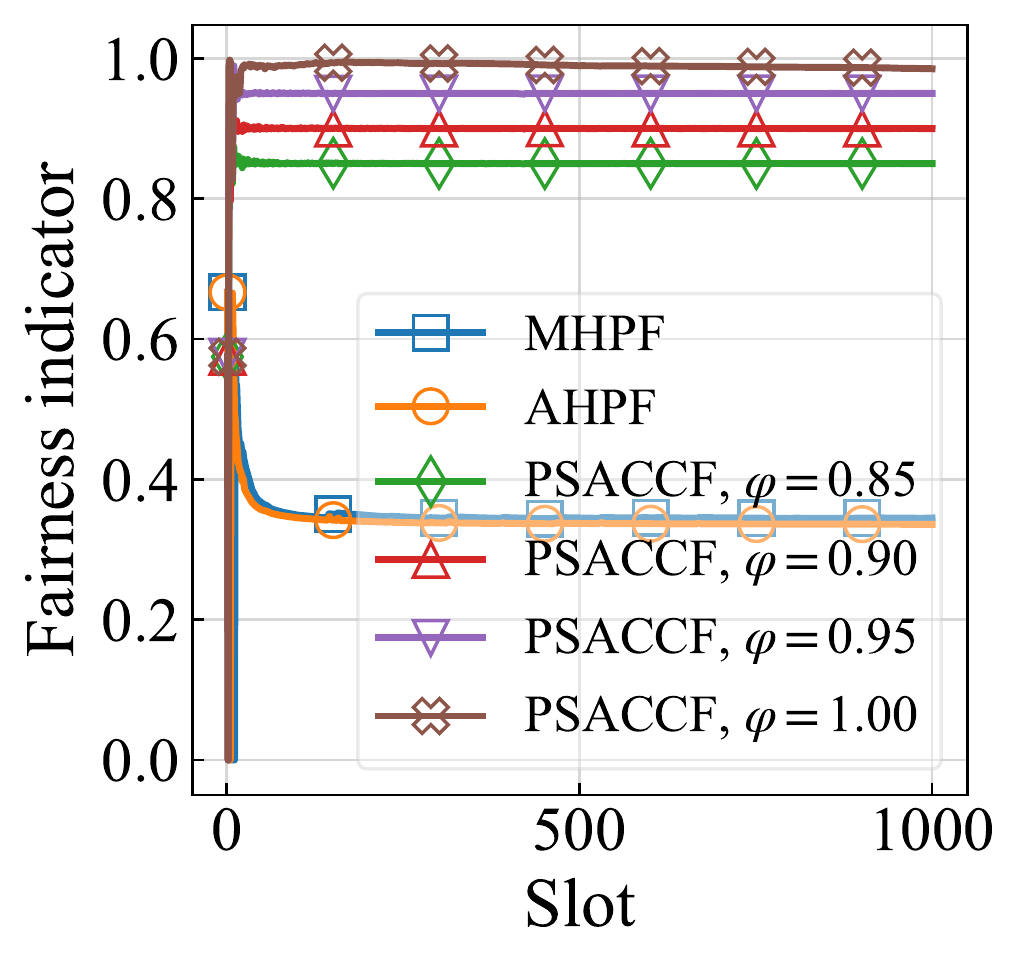}
	}

	\caption{Fairness indicator of three algorithms when $\epsilon=0.15$} \label{fig:Fairness indicator}
\end{figure}

\subsubsection{Fairness indicator} \ \par

Fairness indicator shown in Fig.~\ref{fig:Fairness indicator} exhibits different phenomena from that of priority. The parameter $\epsilon$ is fixed to 0.15 in this graph, and results yielded by PSACCF under all $\varphi$ are drawn together with MHPF and AHPF. Both MHPF and AHPF are not good at reaching a high level of fairness and even get worse when $\lambda_0$ increases due to excessive resources leaning towards services with higher priority. The instability of AHPF is also evident here for the same reason in Fig.~\ref{fig:Priority indicator lambda=5}. However, PSACCF appears to be more robust when facing a changing arrival rate, fairness curves stay roughly at the position where $\varphi$ resides and keep relatively flat. The improvements over MHPF and AHPF are about 33.6\% and 82.9\% respectively when the base arrival rate is 5, and extend to 146.4\% and 153.1\% with $\lambda_0$ tripled.

%Unlike priority, curves of the fairness indicator shown in  Fig.~\ref{fig:Fairness indicator} exhibit obvious differences. PSACCF with fixed $\epsilon=0.15$ but all $\varphi$ are drawn together with MHPF and AHPF in this graph. The figure shows that neither MHPF nor AHPF is very good at achieving a high level of fairness while maintaining priority, both of them obtain poorer fairness when $\lambda_0$ increases because of excessive leaning of resources towards higher priorities, and AHPF even suffers from severe shaking under light loads for the same reason in Fig.~\ref{fig:Priority indicator lambda=5}. PSACCF appears to be more robust in the face of changing arrival rates, and curves are restricted near the positions of $\varphi$ with no much ups and downs overall. 

\begin{figure}[!h]
	\centering
	\subfloat[$\varphi=1$]{
		\centering
		\includegraphics[scale=0.4]{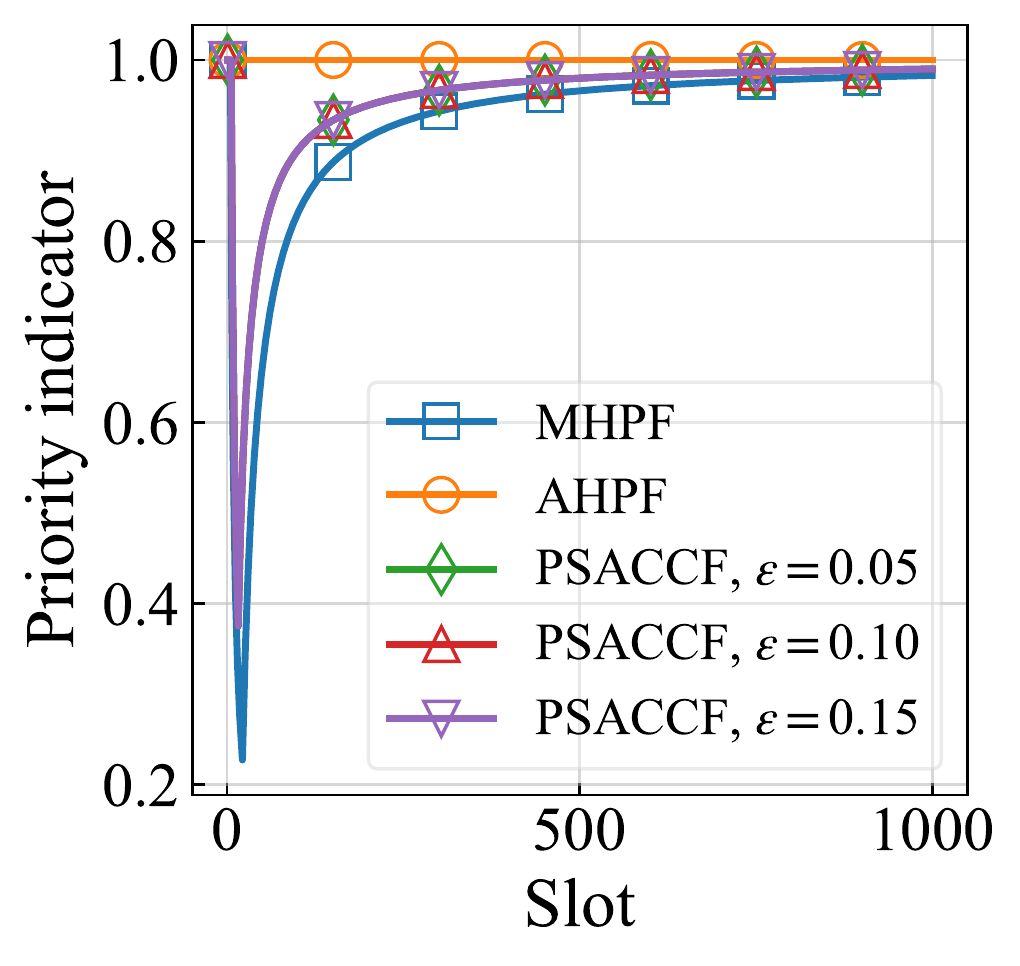}
		\label{fig:priority no balk}
	}
	\subfloat[$\epsilon=0.15$]{
		\centering
		\includegraphics[scale=0.4]{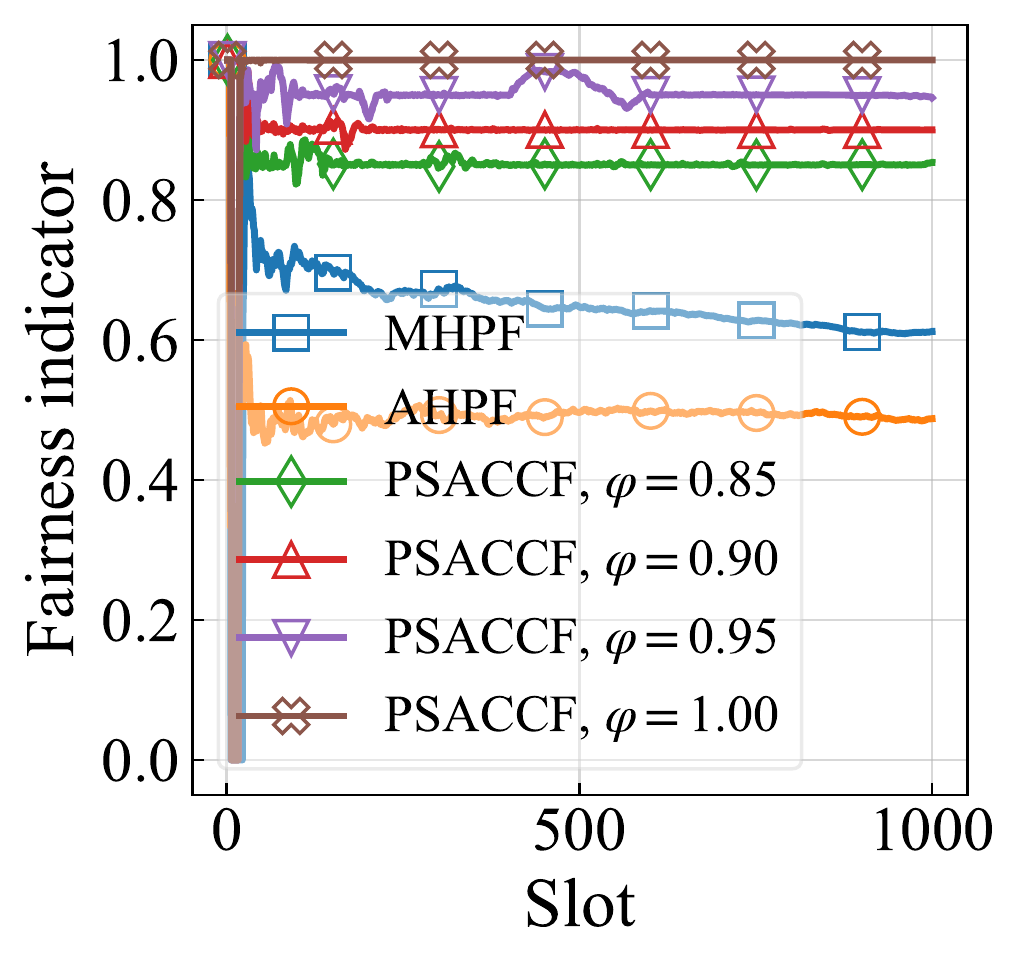}
		\label{fig:fairness no balk}
	}
	
	\caption{Performance indicators with no balk when $\lambda_0=5$} \label{fig:no balk}
\end{figure}

To show that the odd performance of AHPF at low workload is indeed due to balking, we forbid it by setting $\beta_0$ to zero in Fig.~\ref{fig:no balk}. Unsurprisingly, AHPF obtains the best priority performance all the time but is eventually caught up by others. Although there are no jumps in the fairness curve of AHPF, it's still the worst of them, followed by MHPF, and PSACCF best with 38.6\% and 74.1\% enhancement.

It can be inferred from these graphs that our algorithm is highly controllable and extremely versatile due to its sensitivity to endogenous parameters and stability to environmental settings. Pretty fairness can be achieved by pulling $\varphi$ to 1, and one can degrade PSACCF to MHPF by suppressing $\varphi$ to zero when fairness is completely unimportant.
% the SP can simply pull $\varphi$ full when pretty fairness is needed, or degrade it to MHPF by suppressing $\varphi$ to zero when fairness is not cared about at all.

\subsubsection{Resource utilization} \ \par

\begin{figure*}[!tb]
	\centering	
	\subfloat[$\lambda_0=5, \epsilon=0.15, \varphi=0.85$]{
		\centering
		\includegraphics[scale=0.35]{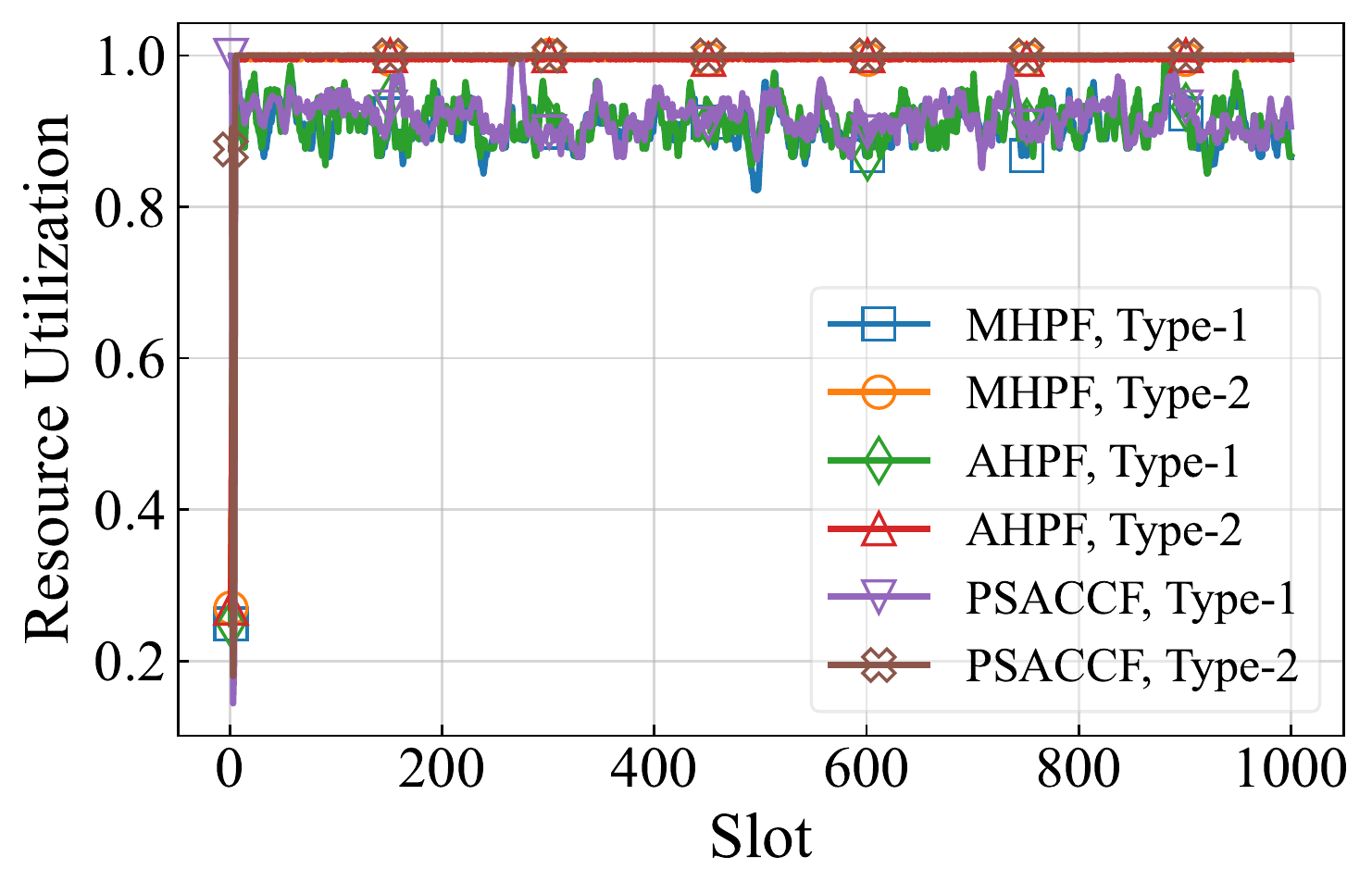}
	}
	\subfloat[$\lambda_0=11, \epsilon=0.15, \varphi=0.85$]{
		\centering
		\includegraphics[scale=0.35]{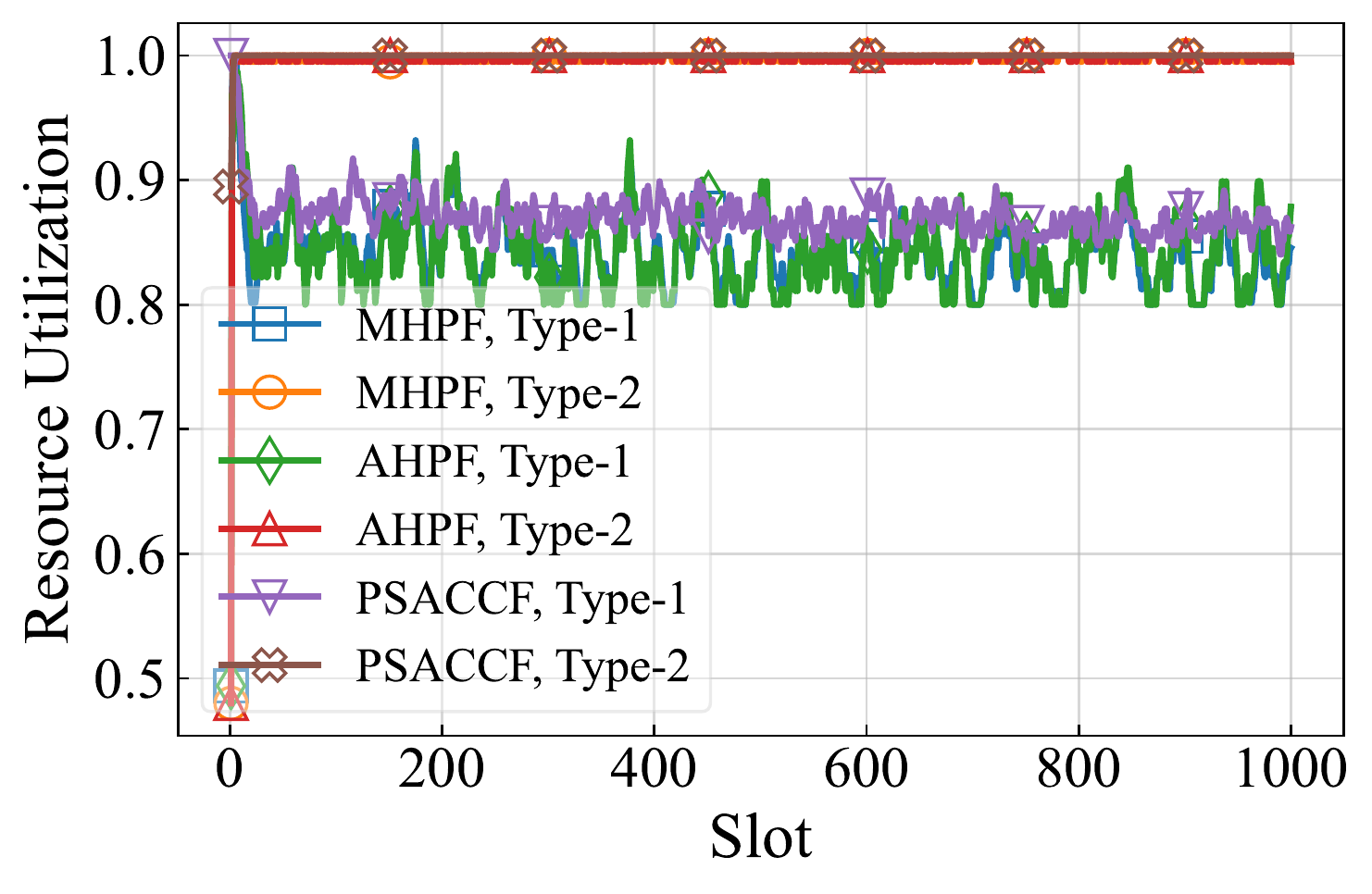}
		\label{fig:ResUti lambda=11 epsilon =0.15 varphi=0.85}
	}
	\subfloat[$\lambda_0=15, \epsilon=0.15, \varphi=0.85$]{
		\centering
		\includegraphics[scale=0.35]{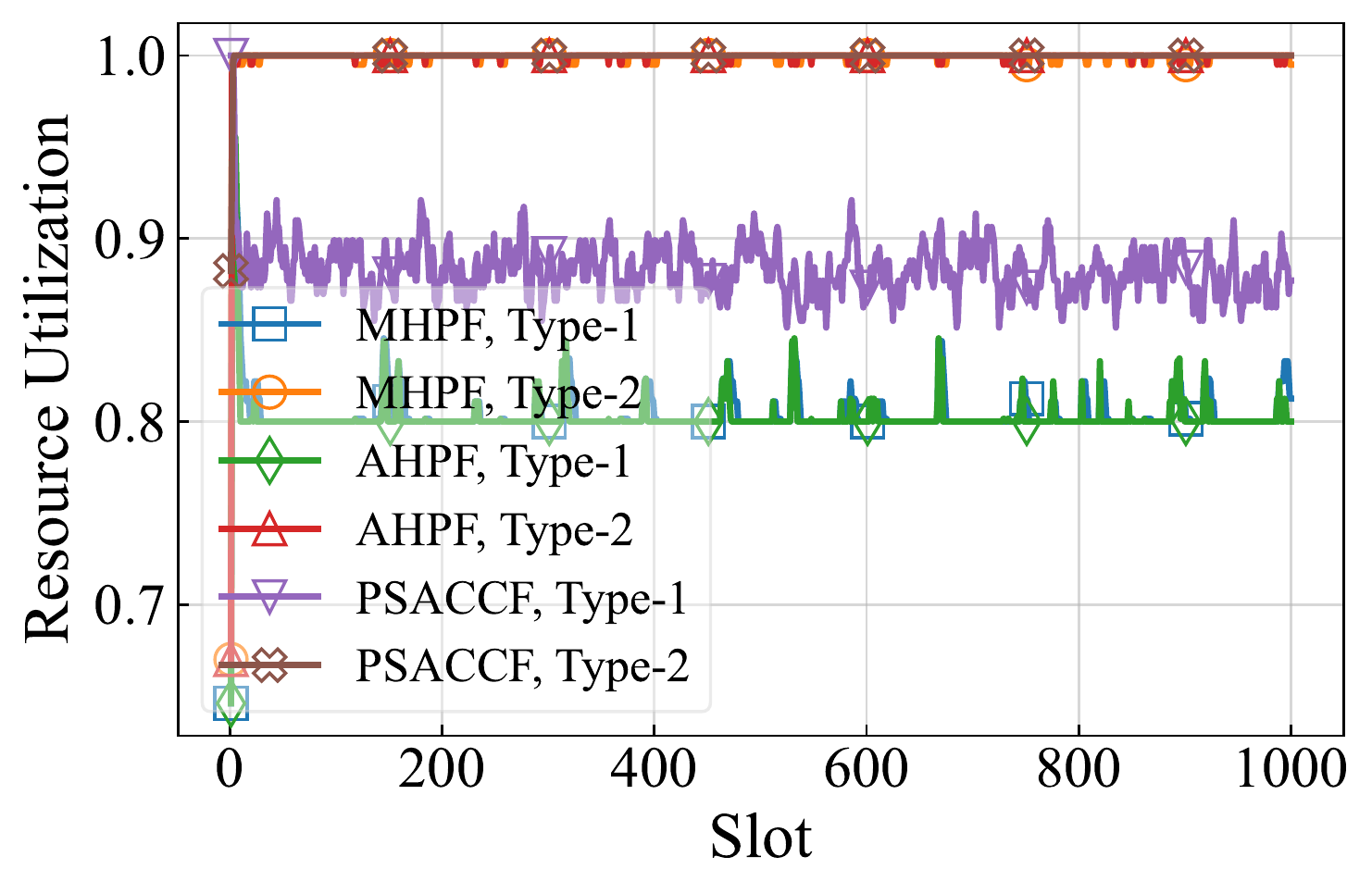}
		\label{fig:ResUti lambda=15 epsilon =0.15 varphi=0.85}
	}
	
	\subfloat[$\lambda_0=5, \epsilon=0.05, \varphi=1$]{
		\centering
		\includegraphics[scale=0.35]{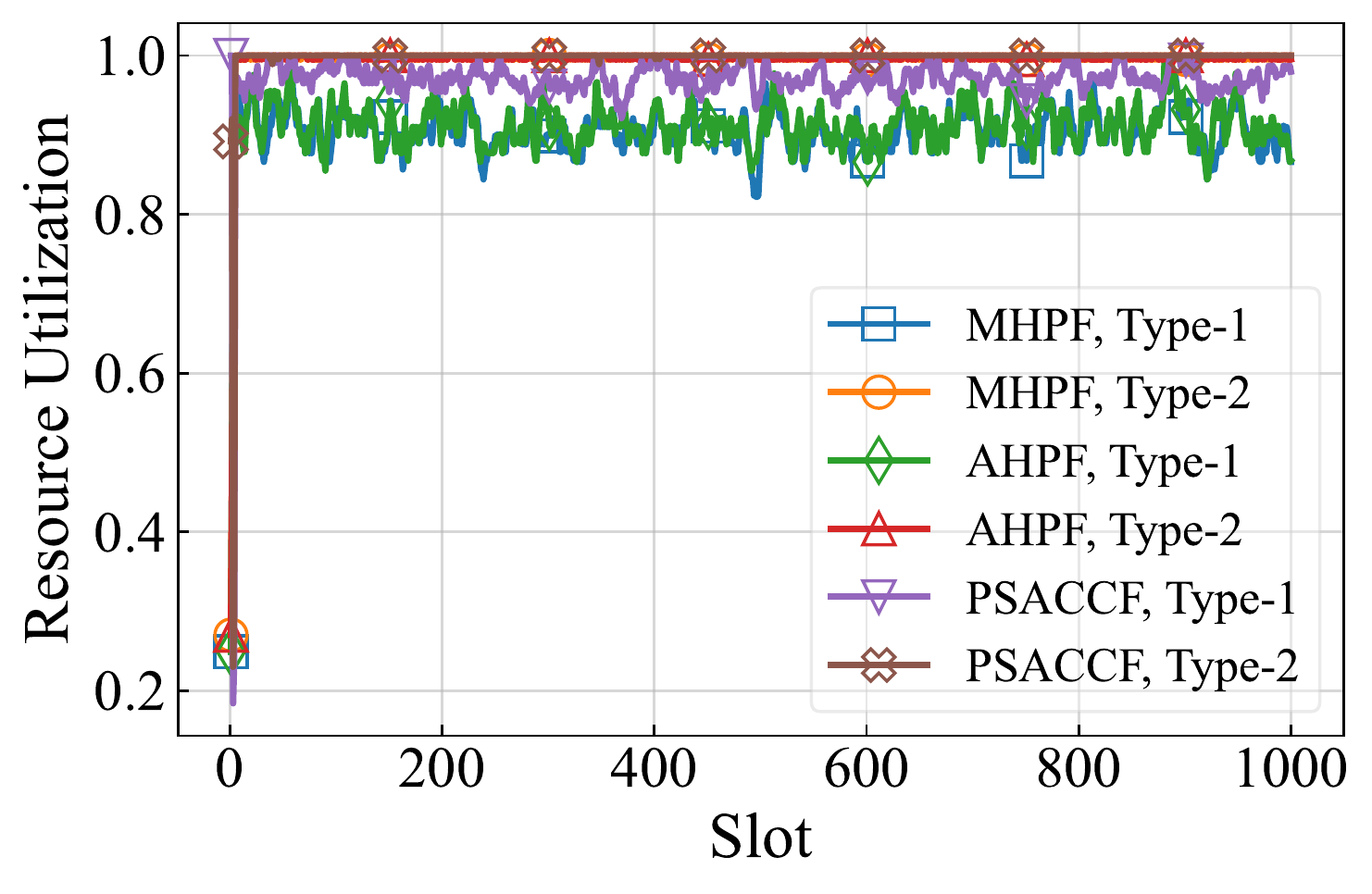}
	}
	\subfloat[$\lambda_0=11, \epsilon=0.05, \varphi=1$]{
		\centering
		\includegraphics[scale=0.35]{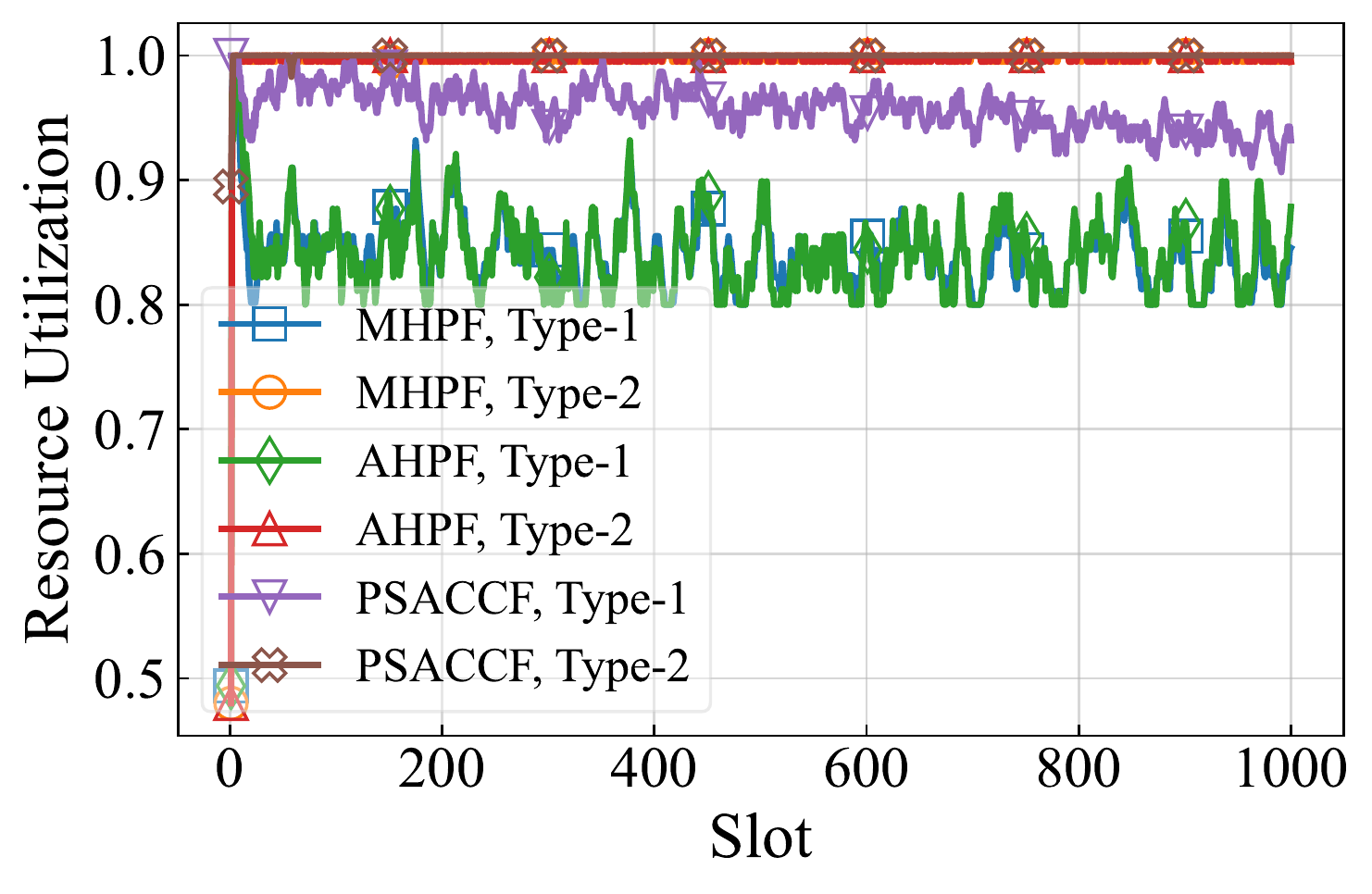}
		\label{fig:ResUti lambda=11 epsilon =0.05 varphi=1}
	}
	\subfloat[$\lambda_0=15, \epsilon=0.05, \varphi=1$]{
		\centering
		\includegraphics[scale=0.35]{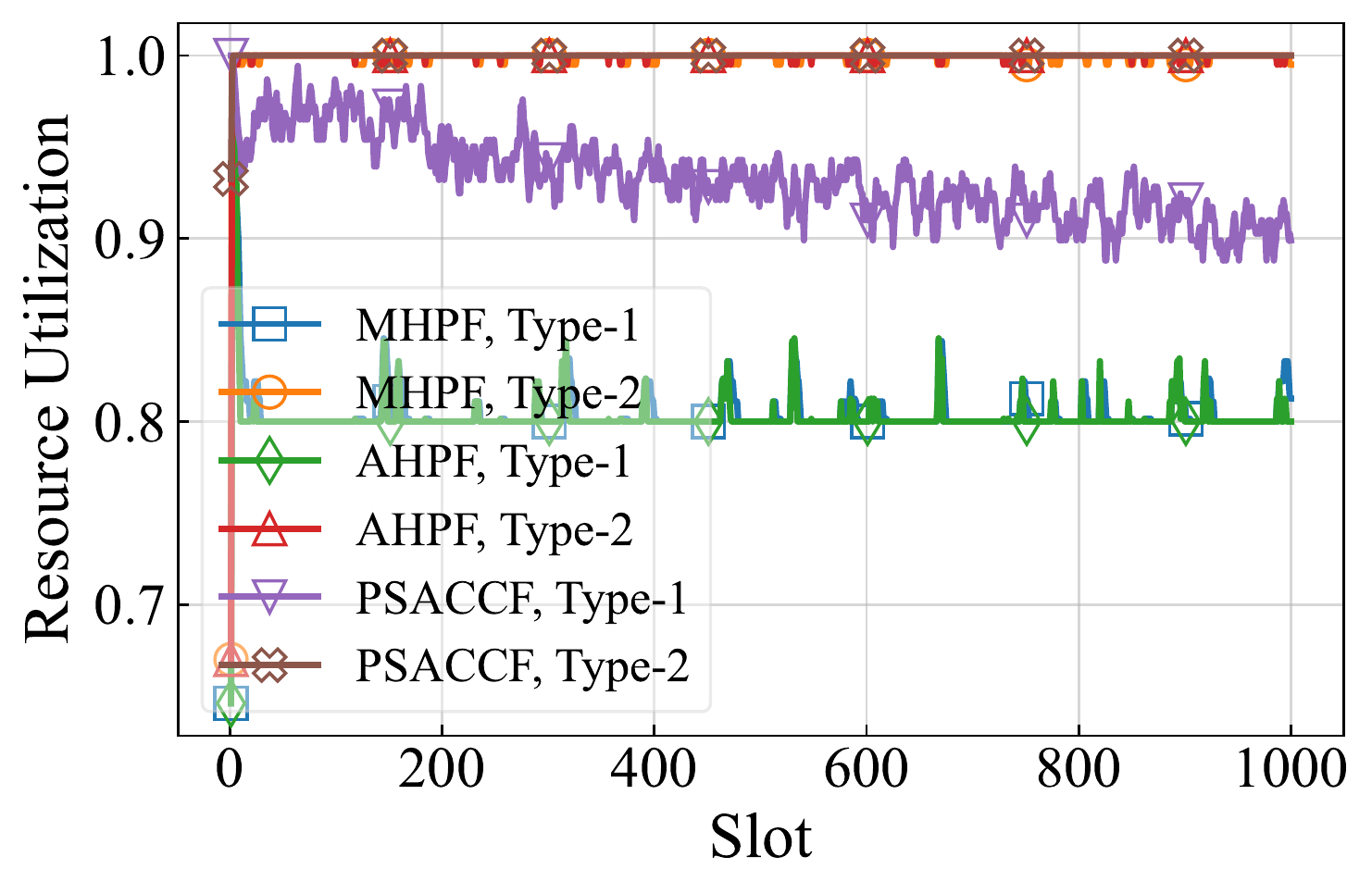} 
		\label{fig:ResUti lambda=15 epsilon =0.05 varphi=1}
	}
	\caption{Resource utilization of three algorithms} \label{fig:ResUti}
\end{figure*}

We further examined the resource utilization of each algorithm in Fig.~\ref{fig:ResUti}. For the Type-2 resource, all of them achieve almost 100\% utilization. In contrast, Type-1 is less utilized and keeps decreasing as $\lambda_0$ increases. It's because of the priority requirement concerned by the three algorithms. Resource of Type-2 is much more dominant than Type-1 in Service-4, which has the highest priority level, so the second resource will be exhausted first, leaving idle resources in Type-1. When the base arrival rate increases, more high-priority requests are accepted, so the utilization of the first resource decreases accordingly.

Although average resource utilization of the Type-1 resource under the three algorithms does not come to 100\%, PSACCF is significantly higher, around 3.4\% and 2.5\% promotions over MHPF and AHPF in Fig.~\ref{fig:ResUti lambda=11 epsilon =0.15 varphi=0.85}, extend to 9.2\% and 9.6\% when $\lambda_0$ gets larger in Fig.\ref{fig:ResUti lambda=15 epsilon =0.15 varphi=0.85}. This is a consequence of fairness optimization. The access opportunities are distributed among requests with better balance by PSACCF, so a portion of resources are reserved for Services-1 and Service-3, whose demand for Type-1 exceeds those for Type-2, so the minimum resource utilization rises. What's more, PSACCF can be parameterized with narrower $\epsilon$ or bigger $\varphi$ to increase the lifting range, which is shown in each column of Fig.~\ref{fig:ResUti}. Promotions in Fig.~\ref{fig:ResUti lambda=11 epsilon =0.05 varphi=1} are 12.4\% and 11.5\%, in Fig.~\ref{fig:ResUti lambda=15 epsilon =0.05 varphi=1} are 13.0\% and 13.4\%.

%Fortunately, range of drop is effectively suppressed in PSACCF, which is precisely due to the consideration of fairness. The opportunities of admission are distributed more balanced among requests by PSACCF, so a portion of resources are reserved for Services 1 and 3, where demands for Type-1 exceed those for Type-2. This is why the minimum resource utilization is optimized. What's more, lower curve of PSACCF can move to a certain position after adjusting the two endogenous parameters, i.e., $\epsilon$ and $\varphi$, which are shown in each column of Fig.~\ref{fig:ResUti}.

\subsubsection{Profit and time consumption} \ \par

For the sake of comprehensiveness of comparison, the profit and time cost of the three algorithms are included in simulations. Fig.~\ref{fig:Profit triple} shows the cumulative profit earned by the SP when a totally 1000 slots have passed. The best curve belongs to MHPF, followed by AHPF when $\lambda_0$ is less than 13. Because resource hogging in AHPF will cause plenty of SLA violations and incur a penalty. On the other hand, when $\lambda_0$ is too high to occur a preemption, the profit of AHPF and MHPF can hardly be distinguished. PSACCF suffers some loss of earnings, which is one unavoidable cost to pursue extreme fairness (i.e., $\varphi$ adopts 1 in this figure). Fortunately, the amount of loss can be effectively controlled with a bigger $\epsilon$, and would be even smaller if the relaxation of fairness is permitted, which has been demonstrated in Fig.~\ref{fig:PSACCF profit varphi}.

%For the sake of data completeness, we also compared the total profit curves of the three algorithms, results are shown in Fig.~\ref{fig:Profit triple}. The best curve belongs to MHPF, followed by AHPF when $\lambda_0$ is less than 13. Because there are penalties for SLA violations caused by resource hogging that prevents some less privileged ongoing requests from being completed on time. Therefore, when $\lambda_0$ is too high to exist preemptions, profits of AHPF and MHPF can hardly be distinguished. PSACCF, on the other hand, has lost some of its gains due to the pursuit of fairness. If fairness has to be considered, the profit is an unavoidable cost for PSACCF, but if fairness is less important, it is absolutely possible to pursue a higher gain by increasing $\epsilon$ and decreasing $\varphi$.

\begin{figure}[!h]
	\centering
	\includegraphics[scale=0.4]{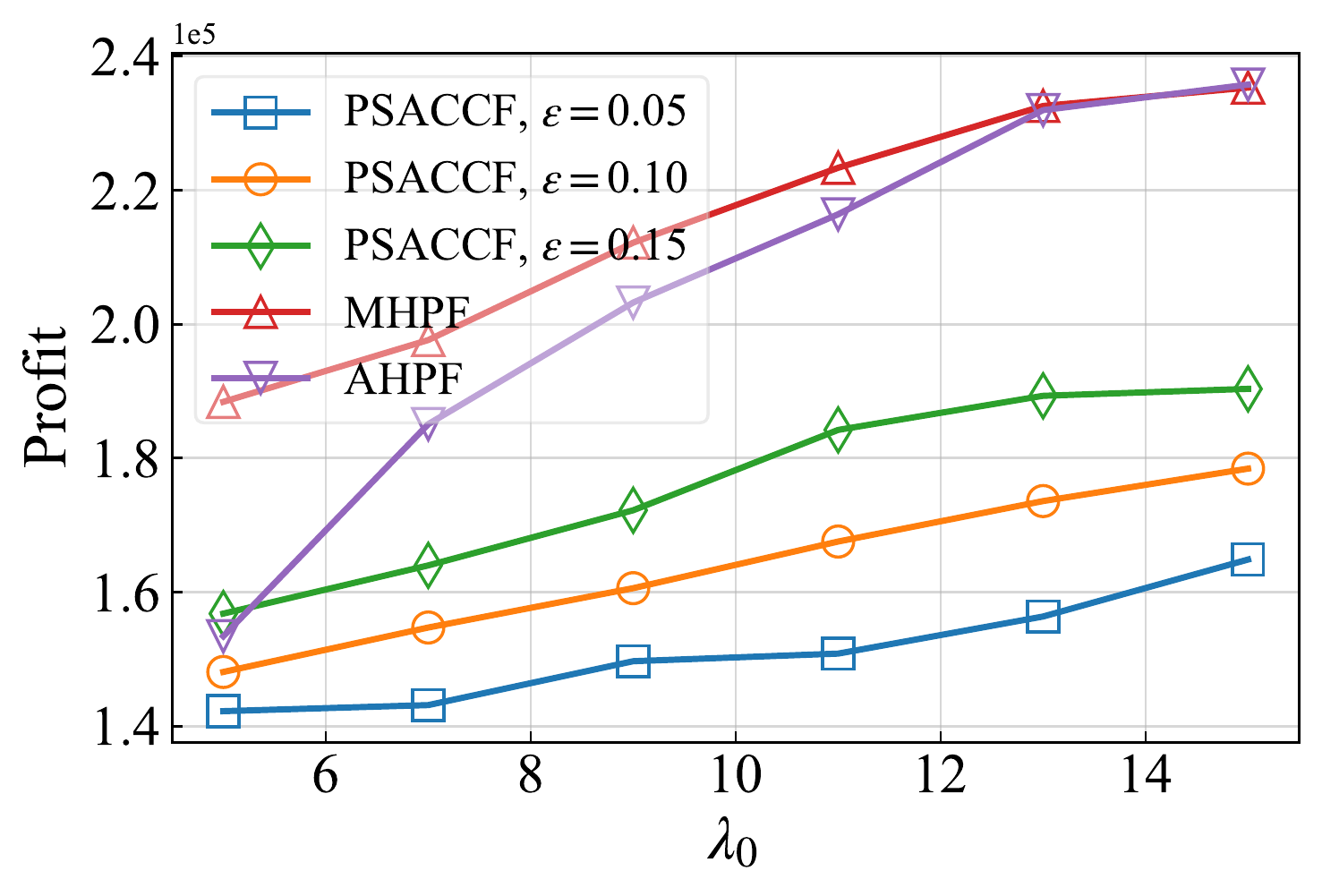}
	\caption{Profit of three algorithms} \label{fig:Profit triple}
\end{figure}

Finally, we provide the time performance in Fig.~\ref{fig:time consumption}. Each curve consists of a concatenation of the total time spent per slot calculating the admission decision and allocating resources. Both MHPF and AHPF are with very low time overhead as they only consider resource availability so that it appears that PSACCF consumes relatively much more time. Actually, the time cost of PSACCF fluctuates mainly between 1ms and 4ms, which is small enough for applications where the time slot usually scales in minutes, seconds, and hundreds of milliseconds. What's more, adjustments of $\epsilon$ and $\varphi$ hardly influence the time cost, ensuring that PSACCF will not suffer from time availability problems when adjusting parameters for different scenarios.

\begin{figure}[!h]
	\centering
	\subfloat[$\epsilon=0.10$]{
		\centering
		\includegraphics[scale=0.35]{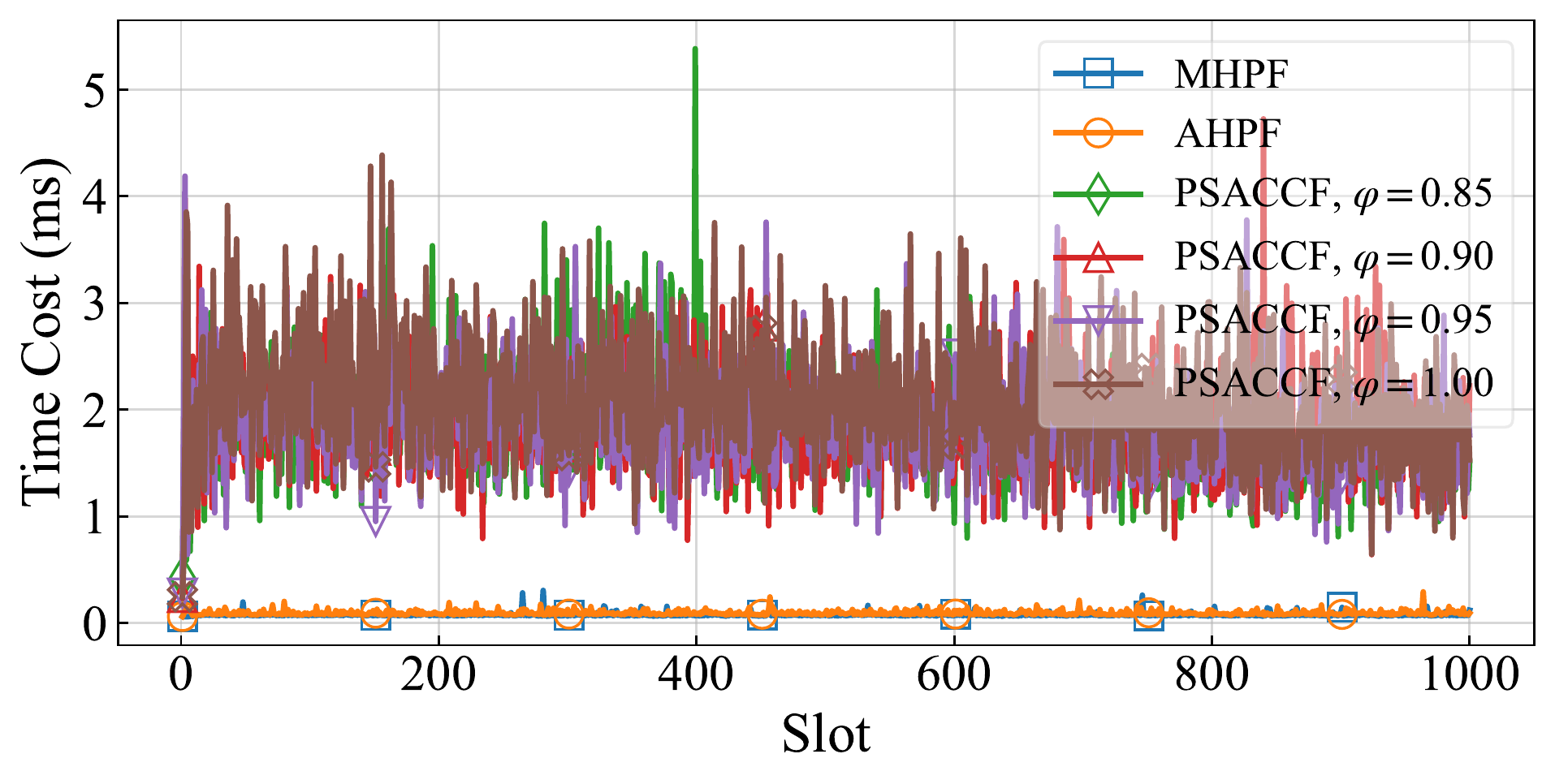} 
	}

	\subfloat[$\varphi=1$]{
	\centering
	\includegraphics[scale=0.35]{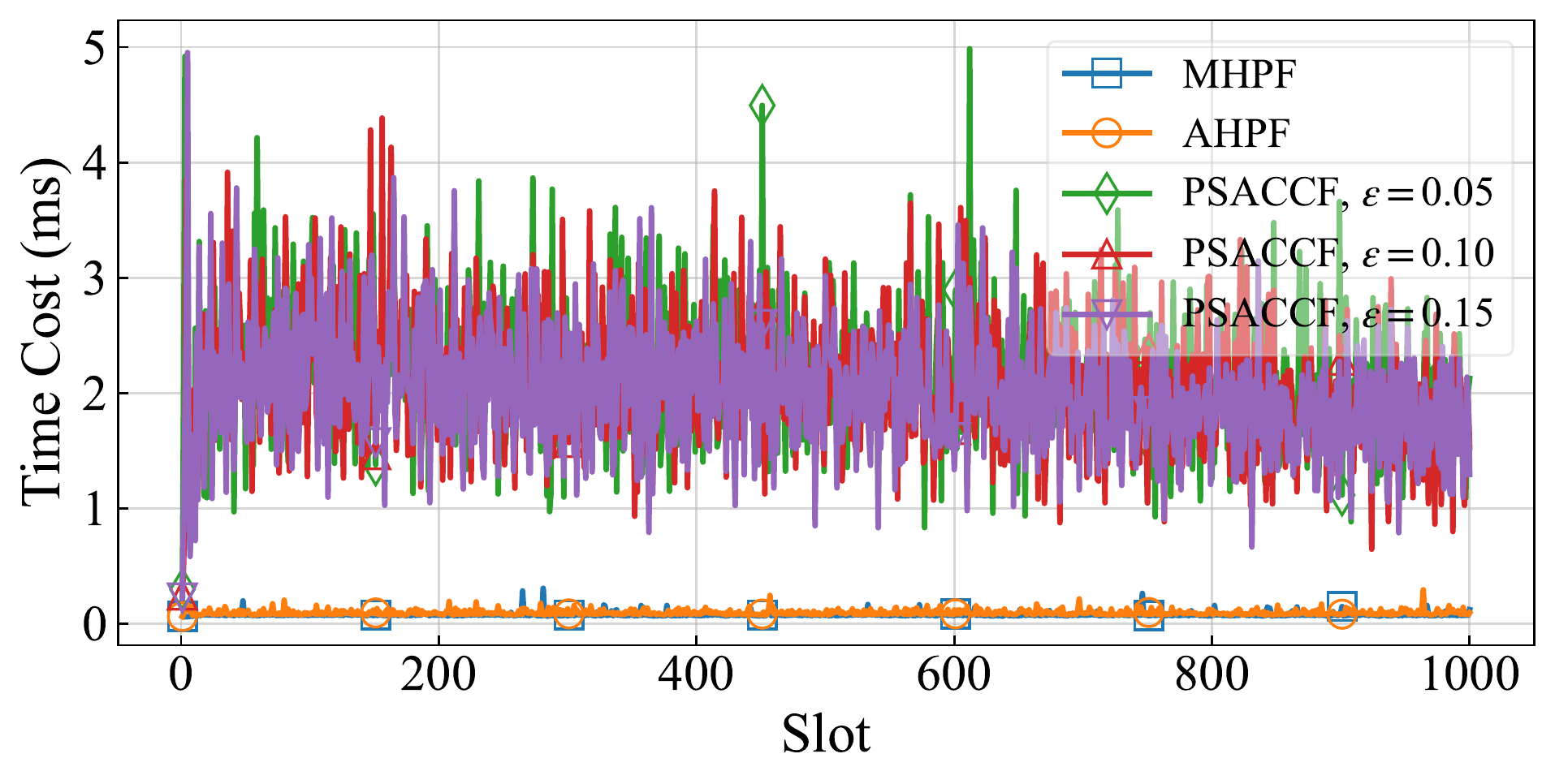} 
	}
	\caption{Time consumption when $\lambda_0=11$} \label{fig:time consumption}
\end{figure}

Overall, these simulation results provide sufficient and convincing proofs that the algorithm proposed in this work can achieve better fairness and resource utilization while ensuring the priority requirement for slice admission control. Despite there exists a certain sacrifice in terms of time costs, the degradation is relatively small compared with the length of time slot, thus does not affect the main functionality and usability of our algorithm. In addition, two adjustable parameters grant PSACCF a great generalization capability to cope with specific requirements in various scenarios.

}

%% file: section/sec07.ConclusionFutureWork.tex
\section{Conclusion and Future Work} \label{conclusion}

In this work, we investigate the slice admission control problem in 5G/B5G networks where different services require heterogeneous resource demands, and rational behaviors of subscribers are considered. The main focus here is to optimize the fairness of admission decisions as much as possible while respecting differences in service priorities.

We began by reinterpreting the concepts of prioritization and fairness in the Slice-as-a-Service paradigm so that they can more accurately describe the business characteristics and requirements in the new scenario and are compatible with each other. Then we proposed a heuristic SAC algorithm called PSACCF based on new definitions to take responsibility for making admission decisions. Simulation results show that PSACCF achieves a comparable priority metric to the comparison algorithms, and obtains a significantly better fairness indicator meanwhile. More noteworthy is PSACCF shows good stability to changes in external parameters, and is gifted in handling various objectives because of the generalizability and adaptability granted by its endogenous parameters.

{\color{black}
In the future, we pay attention to the slice admission control problem with the presence of multiple competing service providers. In that circumstances, each SP may carefully adjust his admission parameters to attract more subscribers from his counterparts for concerns like profit, which can lead to gaming behaviors among SPs. Thus, we believe this will be a much complex and interesting issue for investigations.
}

\section*{Acknowledgments}

This research was partially supported by the National Key Research and Development Program of China (2019YFB1802800), PCL Future Greater-Bay Area Network Facilities for Large-scale Experiments and Applications (LZC0019).

%% file: PSACCF.bbl
\begin{thebibliography}{10}
\providecommand{\url}[1]{#1}
\csname url@samestyle\endcsname
\providecommand{\newblock}{\relax}
\providecommand{\bibinfo}[2]{#2}
\providecommand{\BIBentrySTDinterwordspacing}{\spaceskip=0pt\relax}
\providecommand{\BIBentryALTinterwordstretchfactor}{4}
\providecommand{\BIBentryALTinterwordspacing}{\spaceskip=\fontdimen2\font plus
\BIBentryALTinterwordstretchfactor\fontdimen3\font minus
  \fontdimen4\font\relax}
\providecommand{\BIBforeignlanguage}[2]{{%
\expandafter\ifx\csname l@#1\endcsname\relax
\typeout{** WARNING: IEEEtran.bst: No hyphenation pattern has been}%
\typeout{** loaded for the language `#1'. Using the pattern for}%
\typeout{** the default language instead.}%
\else
\language=\csname l@#1\endcsname
\fi
#2}}
\providecommand{\BIBdecl}{\relax}
\BIBdecl

\bibitem{shafi20175g}
M.~Shafi, A.~F. Molisch, P.~J. Smith\emph{,~et~al.}, ``5g: A tutorial overview
  of standards, trials, challenges, deployment, and practice,'' \emph{IEEE
  Journal on Selected Areas in Communications}, vol.~35, no.~6, pp. 1201--1221,
  2017.

\bibitem{chahbar2020comprehensive}
M.~Chahbar, G.~Diaz, A.~Dandoush\emph{,~et~al.}, ``A comprehensive survey on
  the e2e 5g network slicing model,'' \emph{IEEE Transactions on Network and
  Service Management}, vol.~18, no.~1, pp. 49--62, 2020.

\bibitem{zhang2019overview}
S.~Zhang, ``An overview of network slicing for 5g,'' \emph{IEEE Wireless
  Communications}, vol.~26, no.~3, pp. 111--117, 2019.

\bibitem{posner1985traffic}
E.~C. Posner, ``Traffic policies in cellular radio that minimize blocking of
  handoff calls,'' in \emph{Proc. 11th Teletraffic Congress (ITC 11), vol. 1,
  Kyoto}, vol.~2, 1985.

\bibitem{ramjee1997optimal}
R.~Ramjee, D.~Towsley, and R.~Nagarajan, ``On optimal call admission control in
  cellular networks,'' \emph{Wireless Networks}, vol.~3, no.~1, pp. 29--41,
  1997.

\bibitem{ojijo2020survey}
M.~O. Ojijo and O.~E. Falowo, ``A survey on slice admission control strategies
  and optimization schemes in 5g network,'' \emph{IEEE Access}, vol.~8, pp.
  14\,977--14\,990, 2020.

\bibitem{tang2019service}
J.~Tang, B.~Shim, and T.~Q. Quek, ``Service multiplexing and revenue
  maximization in sliced c-ran incorporated with urllc and multicast embb,''
  \emph{IEEE Journal on Selected Areas in Communications}, vol.~37, no.~4, pp.
  881--895, 2019.

\bibitem{li2021automated}
X.~Li, C.~F. Chiasserini, J.~Mangues-Bafalluy\emph{,~et~al.}, ``Automated
  service provisioning and hierarchical sla management in 5g systems,''
  \emph{IEEE Transactions on Network and Service Management}, 2021.

\bibitem{nourian2021practical}
M.~Nourian, A.~Kusedghi, and A.~Akbari, ``A practical resource management
  prototype for mobile networks,'' in \emph{2021 26th International Computer
  Conference, Computer Society of Iran (CSICC)}.\hskip 1em plus 0.5em minus
  0.4em\relax IEEE, 2021, pp. 1--6.

\bibitem{zhang2020discussion}
Y.~Zhang and Y.~Zhang, ``Discussion on key technologies of cloud game based on
  5g and edge computing,'' in \emph{2020 IEEE 20th International Conference on
  Communication Technology (ICCT)}.\hskip 1em plus 0.5em minus 0.4em\relax
  IEEE, 2020, pp. 524--527.

\bibitem{siriwardhana2021survey}
Y.~Siriwardhana, P.~Porambage, M.~Liyanage\emph{,~et~al.}, ``A survey on mobile
  augmented reality with 5g mobile edge computing: Architectures, applications,
  and technical aspects,'' \emph{IEEE Communications Surveys \& Tutorials},
  vol.~23, no.~2, pp. 1160--1192, 2021.

\bibitem{hwang2005call}
Y.~H. Hwang and S.-K. Noh, ``A call admission control scheme for heterogeneous
  service considering fairness in wireless networks,'' in \emph{Fourth Annual
  ACIS International Conference on Computer and Information Science
  (ICIS'05)}.\hskip 1em plus 0.5em minus 0.4em\relax IEEE, 2005, pp. 688--692.

\bibitem{ogryczak2014fair}
W.~Ogryczak, H.~Luss, M.~Pi{\'o}ro\emph{,~et~al.}, ``Fair optimization and
  networks: A survey,'' \emph{Journal of Applied Mathematics}, vol. 2014, 2014.

\bibitem{bakri2021using}
S.~Bakri, B.~Brik, and A.~Ksentini, ``On using reinforcement learning for
  network slice admission control in 5g: Offline vs. online,''
  \emph{International Journal of Communication Systems}, vol.~34, no.~7, p.
  e4757, 2021.

\bibitem{villota2021admission}
W.~F. Villota~J{\'a}come, O.~M. Caicedo~Rendon, and N.~L.~S. da~Fonseca,
  ``Admission control for 5g network slicing based on (deep) reinforcement
  learning,'' \emph{TechRxiv preprint
  https://doi.org/10.36227/techrxiv.14498190.v1}, 2021.

\bibitem{challa2019network}
R.~Challa, V.~V. Zalyubovskiy, S.~M. Raza\emph{,~et~al.}, ``Network slice
  admission model: Tradeoff between monetization and rejections,'' \emph{IEEE
  Systems Journal}, vol.~14, no.~1, pp. 657--660, 2019.

\bibitem{ebrahimi2020joint}
S.~Ebrahimi, A.~Zakeri, B.~Akbari\emph{,~et~al.}, ``Joint resource and
  admission management for slice-enabled networks,'' in \emph{NOMS 2020-2020
  IEEE/IFIP Network Operations and Management Symposium}.\hskip 1em plus 0.5em
  minus 0.4em\relax IEEE, 2020, pp. 1--7.

\bibitem{han2020multiservice}
B.~Han, V.~Sciancalepore, X.~Costa-Perez\emph{,~et~al.}, ``Multiservice-based
  network slicing orchestration with impatient tenants,'' \emph{IEEE
  Transactions on Wireless Communications}, vol.~19, no.~7, pp. 5010--5024,
  2020.

\bibitem{han2018admission}
B.~Han, A.~DeDomenico, G.~Dandachi\emph{,~et~al.}, ``Admission and congestion
  control for 5g network slicing,'' in \emph{2018 IEEE Conference on Standards
  for Communications and Networking (CSCN)}.\hskip 1em plus 0.5em minus
  0.4em\relax IEEE, 2018, pp. 1--6.

\bibitem{monemi2015low}
M.~Monemi, M.~Rasti, and E.~Hossain, ``Low-complexity sinr feasibility checking
  and joint power and admission control in prioritized multitier cellular
  networks,'' \emph{IEEE Transactions on Wireless Communications}, vol.~15,
  no.~3, pp. 2421--2434, 2015.

\bibitem{caballero2018network}
P.~Caballero, A.~Banchs, G.~De~Veciana\emph{,~et~al.}, ``Network slicing for
  guaranteed rate services: Admission control and resource allocation games,''
  \emph{IEEE Transactions on Wireless Communications}, vol.~17, no.~10, pp.
  6419--6432, 2018.

\bibitem{caballero2019network}
P.~Caballero, A.~Banchs, G.~De~Veciana\emph{,~et~al.}, ``Network slicing games:
  Enabling customization in multi-tenant mobile networks,'' \emph{IEEE/ACM
  Transactions on Networking}, vol.~27, no.~2, pp. 662--675, 2019.

\bibitem{al2020priority}
A.~A. Al-Khatib and A.~Khelil, ``Priority-and reservation-based slicing for
  future vehicular networks,'' in \emph{2020 6th IEEE Conference on Network
  Softwarization (NetSoft)}.\hskip 1em plus 0.5em minus 0.4em\relax IEEE, 2020,
  pp. 36--42.

\bibitem{hwang2003fairness}
R.-H. Hwang and C.-F. Chi, ``Fairness in qos guaranteed networks,'' in
  \emph{IEEE International Conference on Communications, 2003. ICC'03.},
  vol.~1.\hskip 1em plus 0.5em minus 0.4em\relax IEEE, 2003, pp. 218--222.

\bibitem{yang2020genetic}
X.~Yang, Y.~Wang, I.~C. Wong\emph{,~et~al.}, ``Genetic algorithm in resource
  allocation of ran slicing with qos isolation and fairness,'' in \emph{2020
  IEEE Latin-American Conference on Communications (LATINCOM)}.\hskip 1em plus
  0.5em minus 0.4em\relax IEEE, 2020, pp. 1--6.

\bibitem{dianati2005new}
M.~Dianati, X.~Shen, and S.~Naik, ``A new fairness index for radio resource
  allocation in wireless networks,'' in \emph{IEEE Wireless Communications and
  Networking Conference, 2005}, vol.~2.\hskip 1em plus 0.5em minus 0.4em\relax
  IEEE, 2005, pp. 712--717.

\bibitem{caballero2017multi}
P.~Caballero, A.~Banchs, G.~De~Veciana\emph{,~et~al.}, ``Multi-tenant radio
  access network slicing: Statistical multiplexing of spatial loads,''
  \emph{IEEE/ACM Transactions on Networking}, vol.~25, no.~5, pp. 3044--3058,
  2017.

\bibitem{nasser2004optimal}
N.~Nasser and H.~Hassanein, ``An optimal and fair call admission control policy
  for seamless handoff in multimedia wireless networks with qos guarantees,''
  in \emph{IEEE Global Telecommunications Conference, 2004. GLOBECOM'04.},
  vol.~6.\hskip 1em plus 0.5em minus 0.4em\relax IEEE, 2004, pp. 3926--3930.

\bibitem{liu2014fairness}
G.~Liu, M.~Sheng, X.~Wang\emph{,~et~al.}, ``Fairness-based joint call admission
  control for heterogeneous wireless networks: an smdp approach,''
  \emph{Science China Information Sciences}, vol.~57, no.~8, pp. 1--12, 2014.

\bibitem{chiu2005network}
D.~M. Chiu and A.~S. Tam, ``Network fairness for heterogeneous applications,''
  in \emph{Proceedings of ACM SIGCOMM ASIA Workshop}, 2005.

\bibitem{wang2010queueing}
K.~Wang, N.~Li, and Z.~Jiang, ``Queueing system with impatient customers: A
  review,'' in \emph{Proceedings of 2010 IEEE International Conference on
  Service Operations and Logistics, and Informatics}.\hskip 1em plus 0.5em
  minus 0.4em\relax IEEE, 2010, pp. 82--87.

\bibitem{li2021research}
Y.~Li, Y.~Wang, Y.~Jin\emph{,~et~al.}, ``Research on wireless resource
  management and scheduling for 5g network slice,'' in \emph{2021 International
  Wireless Communications and Mobile Computing (IWCMC)}.\hskip 1em plus 0.5em
  minus 0.4em\relax IEEE, 2021, pp. 508--513.

\bibitem{haryadi2017fairness}
S.~Haryadi and D.~R. Aryanti, ``The fairness of resource allocation and its
  impact on the 5g ultra-dense cellular network performance,'' in \emph{2017
  11th International Conference on Telecommunication Systems Services and
  Applications (TSSA)}.\hskip 1em plus 0.5em minus 0.4em\relax IEEE, 2017, pp.
  1--4.

\bibitem{perveen2021dynamic}
A.~Perveen, R.~Abozariba, M.~Patwary\emph{,~et~al.}, ``Dynamic traffic
  forecasting and fuzzy-based optimized admission control in federated 5g-open
  ran networks,'' \emph{Neural Computing and Applications}, pp. 1--19, 2021.

\bibitem{han2019utility}
B.~Han, V.~Sciancalepore, D.~Feng\emph{,~et~al.}, ``A utility-driven
  multi-queue admission control solution for network slicing,'' in \emph{IEEE
  INFOCOM 2019-IEEE Conference on Computer Communications}.\hskip 1em plus
  0.5em minus 0.4em\relax IEEE, 2019, pp. 55--63.

\bibitem{jain1984quantitative}
R.~K. Jain, D.-M.~W. Chiu, W.~R. Hawe\emph{,~et~al.}, ``A quantitative measure
  of fairness and discrimination,'' \emph{Eastern Research Laboratory, Digital
  Equipment Corporation, Hudson, MA}, 1984.

\bibitem{kushchazli2021model}
A.~Kushchazli, A.~Ageeva, I.~Kochetkova\emph{,~et~al.}, ``Model of radio
  admission control for urllc and adaptive bit rate embb in 5g network,''
  \emph{CEUR Workshop Proceedings}, pp. 74--84, 2021.

\end{thebibliography}
